\documentclass[twocolumn,twocolappendix,usenames,dvipsnames]{aastex7}
\usepackage{capt-of}

\newcommand{\torch}{\texttt{Torch}}
\newcommand{\msun}{M_{\odot}}
\newcommand{\tff}{$t_{\mathrm{ff}}$}
\newcommand{\snojet}{\texttt{S-nojet1}}
\newcommand{\snojettwo}{\texttt{S-nojet2}}
\newcommand{\sdef}{\texttt{S-jet1}}
\newcommand{\sdeftwo}{\texttt{S-jet2}}
\newcommand{\sdefthree}{\texttt{S-jet3}}
\newcommand{\sshort}{\texttt{S-short}}
\newcommand{\sslow}{\texttt{S-slow}}
\newcommand{\sfast}{\texttt{S-fast}}
\newcommand{\slo}{\texttt{S-lowres}}
\newcommand{\shi}{\texttt{S-hires}}
\newcommand{\mnojet}{\texttt{M-nojet}}
\newcommand{\mdef}{\texttt{M-jet}}
\newcommand{\mshort}{\texttt{M-short}}

\usepackage{verbatim}
\usepackage{amsmath}
\usepackage{multirow}
\usepackage{relsize}
\usepackage{graphicx}
\usepackage{xcolor}
\usepackage{appendix}

\shorttitle{Jets in \texttt{Torch}}
\shortauthors{Appel et al.}

\begin{document}

\title{Protostellar Jets in Star Cluster Formation and Evolution: I. Implementation and Initial Results}

\author[0000-0002-6593-3800]{Sabrina M. Appel}
\altaffiliation{NSF Astronomy \& Astrophysics Postdoctoral Fellow}
\email{sappel@amnh.org}
\affiliation{Department of Astrophysics,
American Museum of Natural History,
New York, NY, USA}
\affiliation{Department of Physics and Astronomy, 
Rutgers University,
Piscataway, NJ, USA}

\author[0000-0001-5817-5944]{Blakesley Burkhart}
\email{bburkhart@flatironinstitute.org}
\affiliation{Department of Physics and Astronomy, 
Rutgers University,
Piscataway, NJ, USA}
\affiliation{Center for Computational Astrophysics, 
Flatiron Institute, 
New York, NY, USA}

\author[0000-0003-0064-4060]{Mordecai-Mark Mac Low}
\email{mordecai@amnh.org}
\affiliation{Department of Astrophysics,
American Museum of Natural History,
New York, NY, USA}
\affiliation{Department of Physics,
Drexel University,
Philadelphia, PA, USA}

\author[0000-0003-3479-4606]{Eric P. Andersson}
\email{eandersson@amnh.org}
\affiliation{Department of Astrophysics,
American Museum of Natural History,
New York, NY, USA}

\author[0000-0002-6116-1014]{Claude Cournoyer-Cloutier}
\email{cournoyc@mcmaster.ca}
\affiliation{Department of Physics and Astronomy,
McMaster University,
Hamilton, ON, Canada}

\author[0000-0003-4866-9136]{Sean Lewis}
\email{sean.phys@gmail.com}
\affiliation{Department of Physics,
Drexel University,
Philadelphia, PA, USA}

\author[0000-0001-9104-9675]{Stephen L. W. McMillan}
\email{slm23@drexel.edu}
\affiliation{Department of Physics,
Drexel University,
Philadelphia, PA, USA}

\author[0000-0001-5972-137X]{Brooke Polak}
\email{bpolak@amnh.org}
\affiliation{Universit\"at Heidelberg, 
Zentrum f\"ur Astronomie,
Institut f\"ur Theoretische Astrophysik,
Heidelberg, Germany}
\affiliation{Department of Astrophysics,
American Museum of Natural History,
New York, NY, USA}

\author[0000-0001-5839-0302]{Simon Portegies Zwart}
\email{spz@strw.leidenuniv.nl}
\affiliation{Leiden Observatory, 
Leiden University, 
Leiden, the Netherlands}

\author[0000-0003-3483-4890]{Aaron Tran}
\email{atran@physics.wisc.edu}
\affiliation{Department of Physics,
University of Wisconsin--Madison,
Madison, WI, USA}

\author[0000-0002-3001-9461]{Maite J. C. Wilhelm}
\email{wilhelm@strw.leidenuniv.nl}
\affiliation{Leiden Observatory, 
Leiden University, 
Leiden, the Netherlands}

\begin{abstract}

Stars form in clusters from the gravitational collapse of giant molecular clouds, which is opposed by a variety of physical processes, including stellar feedback. The interplay between these processes determines the star formation rate of the clouds.
To study how feedback controls star formation, we use a numerical framework that is optimized to simulate star cluster formation and evolution. 
This framework, called \torch, combines the magnetohydrodynamical code FLASH with N-body and stellar evolution codes in the Astrophysical Multipurpose Software Environment (AMUSE).
\torch\ includes stellar feedback from ionizing and non-ionizing radiation, stellar winds, and supernovae, but, until now, did not include protostellar jets.
We present our implementation of protostellar jet feedback within the \torch\ framework and describe its free parameters.
We then demonstrate our new module by comparing cluster formation simulations with and without jets.
We find that the inclusion of protostellar jets slows star formation, even in clouds of up to $M = 2 \times 10^4~\msun$.
We also find that the star formation rate of our lower mass clouds ($M = 5 \times 10^3~\msun$) is strongly affected by both the inclusion of protostellar jets and the chosen jet parameters, including the jet lifetime and injection velocity.
We follow the energy budget for each simulation and find that the inclusion of jets systematically increases the kinetic energy of the gas at early times.
The implementation of protostellar jet feedback in \torch\ opens new areas of investigation regarding the role of feedback in star cluster formation and evolution.

\end{abstract}


\section{Introduction} \label{sec:intro}

Nearly all star formation takes place in clustered environments \citep{LadaLada2003,PortegiesZwart+2010}.
Many star clusters are subsequently disrupted, forming moving groups or going on to build up the population of field stars \citep[e.g.,][]{LadaLada2003, Gagne+2021}.
Thus, studying each stage of the formation, evolution, and potential disruption of clusters is critical for fully understanding star formation, the properties of star clusters, and the population of Galactic field stars. 
Detailed simulations of the formation and evolution of star clusters can give us insights into the roles played by various physical processes during star formation.

At each stage of the star formation process, stars produce various modes of stellar feedback.
Previous work has shown that this stellar feedback plays an important role in shaping how star formation proceeds, including setting the star formation rate (SFR), the star formation efficiency (SFE), and the gas and cluster properties \citep[see e.g.,][]{Federrath2015, Bally2016, Krumholz+2019, RosenKrumholz2020, Appel+2022, Lewis+2023, Appel+2023, Andersson2024, Lebreuilly2024}.

\subsection{Stellar Feedback}

As soon as the first stars form from the dense gas of a molecular cloud stellar feedback can have a significant impact on the gas, the stellar population, and how star formation proceeds \citep[e.g.,][]{Appel+2022,Lewis+2023,Andersson2024}.
Radiative feedback heats and ionizes the gas, which slows star formation by preventing collapse \citep[e.g.,][]{Price2009, RosenKrumholz2020,Yaghoobi2023}, alters the initial mass function \citep[IMF; e.g.,][]{Guszejnov+2016, Menon2024}, and shapes the properties of the dense gas of the ISM \citep[e.g.,][]{Menon+2021}.
Stellar winds can have a dramatic effect on the surrounding ISM as they inject high-velocity, hot gas around the star \citep[see][for a review]{Lancaster2024}.
Indeed, \cite{Lewis+2023} show that the winds and radiation from high-mass stars significantly impact the subsequent formation and evolution of star clusters, especially if they form early in the star formation process.

At later times, supernova feedback can strip all the gas from a star-forming region, bringing an end to star formation in that region.
Supernovae may also act to trigger star formation in neighboring regions by compressing the gas \citep{Herrington+2023}.

Protostellar jets are expected to have a much smaller impact at larger scales (e.g., at the scales needed to disperse an entire giant molecular cloud) than other modes of feedback, such as stellar winds, and instead are expected to dominate only at smaller scales or in association with lower-mass stars \citep{Chevance+2023}.
However, observations find that protostellar jets and outflows are produced by essentially all newly forming stars \citep{Shepherd+1996, Richer2000, Beuther2002, Shepherd2003, Duarte-Cabral2013}, although many aspects of this process are still unclear \citep[see][for reviews of protostellar outflows]{Arce+2007, Frank2014, Bally2016}.
Furthermore, since jets are produced at the earliest stages of star formation by newly forming stars, they shape star formation from its earliest stages, including dispersing individual star-forming cores \citep{MatznerMcKee2000}.
Indeed, both observations and simulations have shown that jets significantly impact the density distribution and dynamics of the gas within the star-forming cloud and in the surrounding ISM \citep{Shepherd2003, Nakamura2007,Federrath2015, Kavak+2022, Appel+2022, Appel+2023, Bally2024}; reduce the SFE \citep{Federrath2014, Krumholz2014, Krumholz+2014, Federrath2015, Bally2016, RosenKrumholz2020, Appel+2022, Lebreuilly2024}; and are important in setting the IMF \citep{Krumholz2005, Federrath2014,Krumholz+2014, Kuiper2015, Staff2019, Mathew2021, Guszejnov2022, Lebreuilly2024}.
Indeed, these works suggest that protostellar jets and outflows are an important part of understanding observed features of star formation.

Each of these feedback mechanisms impact the evolution of the star-forming region, and different feedback mechanisms can interact and synergize in various ways. 
Through star cluster simulations, we can disentangle and explore the effects of each of these different feedback mechanisms.
In this paper, we focus on the impacts of protostellar jets.

\subsection{Protostellar Jets} \label{sec:intro-jets}

Protostellar jets and outflows are launched by newly forming stars, as the magnetic field is wound up by the accretion disk \citep[e.g.,][]{Shu+1988,Pelletier+1992,bontemps96,LyndenBell2003, Banerjee2006, Kolligan2018,Rosen+2020}.
Observations \citep[e.g.,][]{GuethGuilloteau1999} and models \citep[e.g.,][]{Lee2001} suggest that a central jet entrains material within a bow-shock to form molecular outflows.
Protostellar jets and outflows, then, have lifetimes corresponding to the length of the main accretion phase, and may last as long as 200 to 500~kyr \citep{Frank2014, Bally2016}.
Protostars of all masses are expected to produce protostellar outflows \citep[see, e.g.,][and references therein]{Shepherd+1996,Richer2000,Beuther2002, Shepherd2003,Wu2004,Duarte-Cabral2013,Bally2016, Chauhan2024}.

Protostellar jets are reported to have velocities ranging from tens to a few hundreds of kilometers per second \citep{Frank2014, Bally2016, Djupvik2016, Wu+2023, Martinez-Henares+2023, Reipurth2023, Rubinstein2023a, Bally2024}, although the velocities evolve significantly with protostellar mass and age \citep[see, e.g.,][]{Lee2020, Chauhan2024}.
For instance, \cite{Bally2024} reports that the fastest outflow components from massive protostars (e.g., a $\sim20~\msun$ protostar) may reach nearly 2,000~km~s$^{-1}$.

Observations find that protostellar jets and outflows can extend 0.1--10~pc or more into the surrounding ISM \citep[see, e.g.,][]{Frank2014, Kong+2019, Lee2020, Reipurth2023, Bally2024, Chauhan2024}.
However, \cite{Frank2014} suggest that, given jet velocities of 100--300~km~s$^{-1}$ and approximate lifetimes of 200~kyr, protostellar jets may be expected to extend as much as 20--60~pc.
There is observational evidence \citep[e.g.,][]{Shepherd+1996} that protostars have different timescales for their outflows.
Since protostellar jets are associated with the accretion phase of the star, they will no longer be driven by a main sequence star.

The rate at which mass is injected by the protostellar jet is tied to the accretion rate of the protostar \citep[e.g.,][]{Bally2016, Nisini2018}.
Theoretical and observational estimates of the jet mass injection rates can vary from $\sim$0.01--0.5 times the accretion rate \citep[see, e.g.,][]{Beuther2002, Banerjee2006, Price2012, Ellerbroek2013, Federrath2014, Bally2016, Nisini2018}.
The mass injection rate may depend on the mass of the protostar and can vary throughout a protostar's lifetime due to a varying accretion rate.
Indeed, many protostellar jets exhibit knots along their length, likely due to a varying mass injection rate \citep{Reipurth2002, Raga2002, Bonito2010, Bally2016, Djupvik2016, Rubinstein2023a, Chauhan2024, Omura2024}.

Observations vary on whether protostellar jets and outflows are found to be perpendicular to gas filaments or to have random orientations.
\cite{Kong+2019} find that protostellar outflows are typically perpendicular to the filament from which the protostar formed.
However, \cite{MakinFroebrich2018} find a distribution of outflow orientations that is in agreement with random jet orientations.
Furthermore, a variety of observations have found evidence of asymmetric outflows, including unipolar outflows, multiple outflows, or precessing jets \citep{Shepherd+1996,Takaishi2024, Sai2024}.
Multiple outflows from a single protostar are likely due to changes in the outflow orientation.
Such changes in the jet injection direction may be due to the presence of a binary companion, an anisotropic accretion event, or turbulence in the dense core \citep{Shepherd+1996, Sai2024}.

\subsection{The \torch\ framework}

\torch~is a numerical framework optimized for simulating the formation and evolution of stellar clusters and includes a variety of stellar feedback mechanisms \citep{Wall+2019, Wall+2020}.
Previous work with \torch~has explored a variety of topics related to star cluster formation, including the impact of early forming massive stars \citep{Lewis+2023}, the effects of binary formation in clusters \citep{Cournoyer-Cloutier2021,Cournoyer-Cloutier2024b, Cournoyer-Cloutier2025}, star formation in massive clusters \citep{Polak2024a}, and the formation of runaway stars \citep{Polak2024b}.
However, none of the work with \torch~so far has included protostellar jets as a mode of stellar feedback, despite the demonstrated importance of protostellar jets in star formation \citep[e.g.,][]{Federrath2015, Appel+2022, Appel+2023, Guszejnov2022, Bally2024}.

Thus, in this paper, we present a new module within the \torch~framework that implements protostellar jets as an additional feedback mechanism.
We present results from several simulations that use the new jets module to explore the role of protostellar jets in the formation of stellar clusters.
The new module promises significant improvements in our ability to simulate small-scale stellar feedback.

The paper is organized as follows: Section~\ref{sec:torch} gives a brief overview of the \torch~code; Section~\ref{sec:jets_module} describes the new module for implementing protostellar jets; Section~\ref{sec:testing} describes the simulation suite we used to demonstrate our new module and to study the effects of protostellar jets; Section~\ref{sec:results} presents the results from these simulations; Section~\ref{sec:discussion} discusses our results, the features of our new jets module, and future work; and we end with our conclusions in Section~\ref{sec:conclusions}. An earlier version of this paper appeared as Chapter 4 of \citet{Appel2024}.

\section{Overview of Torch} \label{sec:torch}

\verb|Torch| is a numerical framework optimized for tracking the formation and evolution of star clusters \citep{Wall+2019, Wall+2020} by coupling the magnetohydrodynamical (MHD) code \verb|FLASH| \citep{Fryxell2000} to the Astrophysical MUlti-purpose Software Environment (\verb|AMUSE|) framework \citep{PortegiesZwart2009,Pelupessy+2013, PortegiesZwart2013, PortegiesZwartMcMillan2018}.
\verb|Torch| uses \verb|FLASH| to simulate the behavior and evolution of gas using adaptive mesh refinement (AMR) MHD solvers and, in the version we use here, \verb|SeBa| to model stellar evolution \citep{PortegiesZwartVerbunt1996, Toonen+2012} and \verb|ph4| to model N-body dynamics \citep{McMillan+2012, PortegiesZwartMcMillan2018}.
\torch~incorporates a number of additional enhancements to the \verb|Flash| code, including star formation, heating and cooling, ionization, ray-tracing \citep{Baczynski2015}, and stellar feedback \footnote{For further discussion of the physics included in the \torch~code see \citet{Wall+2019} for a description of the star formation routine and \citet{Wall+2020} for more details on the feedback implementation.}.
The combination of these codes enables \verb|Torch| to self-consistently model the formation and evolution of stars, binary systems, and stellar clusters in gas-rich environments.

The \torch~code is available on bitbucket\footnote{The main version of \torch~is available here: \url{https://bitbucket.org/torch-sf/Torch/src/main/}} and is undergoing continuous development.
The version that includes jets, as described in Section~\ref{sec:jets_module}, can currently be found in its own branch\footnote{The jets branch can be found here, until its planned integration into the main branch: \url{https://bitbucket.org/torch-sf/torch/src/jets-sabrina/}}.
The specific version of \torch~initially used for the runs presented in Section~\ref{sec:testing} is tagged as \verb|jets-v1.0|.
During the runs, small changes were made to resolve minor bugs in the code without changing the physics implementation.\footnote{ 
At various points, runs progressed using the following commits: \href{https://bitbucket.org/torch-sf/torch/commits/0b433603c13d05dfe3060f01d029bc6f1e808080}{0b43360}, \href{https://bitbucket.org/torch-sf/torch/commits/3276bce9910353b32c139b4601638bf543bf9130}{3276bce}, \href{https://bitbucket.org/torch-sf/torch/commits/6bef1cd240665f22a5cbc53a3a5510def49abe35}{6bef1cd}, and \href{https://bitbucket.org/torch-sf/torch/commits/b362738547a1e2ba5e9d4c6640b502ea3ff250af}{b362738}.}
The final commit of the version of the jets branch that was used for the production runs presented here is tagged as \verb|jets-v1.1| and can be found at \href{https://bitbucket.org/torch-sf/torch/commits/1a1b67d31930485d91f569d09604150959eb432e}{1a1b67d}.
The currently most up to date version of the jets module, is tagged as \verb|jets-v1.2| and can be found at \href{https://bitbucket.org/torch-sf/torch/commits/1aceb49afc8e332264100b0ca51b361774fac294}{1aceb49} (see Appendix~\ref{sec:dtheta}).

\torch~can use the variety of MHD solvers --- with or without AMR --- available through \verb|FLASH|.
The standard \torch~setup currently uses \verb|FLASH| v4.6.2 \citep{Fryxell2000} with the unsplit MHD solver module \citep{Lee2013} including an HLLD Riemann solver \citep{Miyoshi2005} and third-order piecewise parabolic reconstruction \citep{Colella1984}, as well as a multigrid gravity solver \citep{Ricker2008}.
The runs presented here refine on both pressure and temperature using the \texttt{PARAMESH} refinement criterion \citep{MacNeice2000}, which is adapted from the \cite{Lohner1987} estimator and uses a modified second derivative.
These runs also refine on the Jeans length; we require the Jeans length to be resolved by at least 12 cells \citep{Truelove97a, Heitsch2001, Federrath2010a}.
\torch~handles the gravitational interplay between the gas and the particles via a gravity bridge that effectively implements leapfrog integration \citep{fujii2007,Wall+2019}.
We handle radiation through \verb|FLASH| using the radiative transfer code \verb|FERVENT| \citep{Baczynski2015}, which tracks ionizing and non-ionizing feedback.
As described in \cite{Wall+2020}, \torch\ includes heating and ionization from stars, cosmic ray heating, and cooling from atomic gas, molecular gas, and dust grains.
Our runs assume solar metallicity.

In the simulations presented here, \verb|Torch| uses \verb|AMUSE| to link \verb|FLASH| to \verb|ph4|, which handles the N-body dynamics \citep{McMillan+2012}, and \verb|SeBa|, which handles stellar evolution \citep{PortegiesZwartVerbunt1996, Toonen+2012}.
For these simulations \verb|ph4| uses a 15~R$_{\odot}$ softening radius.
See \cite{Wall+2019} for a more detailed discussion of the integration of \texttt{AMUSE} modules into \torch.

\subsection{Star Formation} \label{sec:star_formation}

\verb|Torch| handles star formation with a combination of sink particles and star particles \citep[see][]{Wall+2019}.
A sink particle is formed when gas within a cell exceeds the sink formation density threshold ($\rho_{\mathrm{sink}}$) and meets the criteria for sink formation described in \cite{Federrath2010a}.
In particular, these criteria require that any gas that is in a cell within the sink radius ($R_{\mathrm{sink}}$) of the new sink particle has a central gravitational minimum and is bound, converging, Jeans-unstable, not within the sink radius of another sink particle, and maximally refined in order to form a sink particle.
After the sink is formed, any gas in cells within a sink radius of the sink particle that exceeds the sink threshold density and that is determined to be bound to the sink is accreted onto the sink particle  \citep{Federrath2010a}.
In order to avoid artificial fragmentation, the Jeans length should be resolved with at least four grid cells \citep{Truelove97a}; thus, the sink radius is set to be $R_{\mathrm{sink}} = 2.5 \Delta x_{\mathrm{min}}$, where $\Delta x_{\mathrm{min}}$ is the minimum cell size.
The sink formation threshold density then is also chosen based on the Jeans length:
\begin{equation}
    \rho_{\mathrm{sink}} = \frac{\pi c_{s}^{2}}{4 \, G  R_{\mathrm{sink}}^2} \ ,
    \label{eq:rho_sink}
\end{equation}
where $G$ is the gravitational constant, and $c_{s}$ is the sound speed evaluated using the initial temperature of the gas \citep{Federrath2010a,Polak2024a}.

When each sink particle is formed, it is assigned a new list of stellar masses that Poisson samples a \citet{Kroupa2002} initial mass function (IMF) between $0.08~\msun$ and $150~\msun$. 
When the sink particle has gained an amount of mass that is equal to or greater than the next stellar mass in its list, a new star particle is created with that mass, and that mass is then removed from the sink.
The newly formed star particle is initialized as a zero age main sequence star.
Each new star particle is set to have a position within the radius of the sink particle that spawns it and the velocity of that sink particle, but with a random perturbation to the position and velocity.
See \cite{Wall+2019} for a more detailed description of the star formation prescription.

\subsection{Stellar Feedback}

\torch\ implements radiative feedback including UV radiation pressure, photoelectric heating, photoionization, supernovae, and stellar winds \citep{Wall+2020}.
Radiation, wind, and supernova feedback are only implemented for star particles with mass above corresponding mass thresholds.
For the runs presented here, these modes of feedback were implemented for star particles with masses $>7~\msun$.
See \cite{Wall+2020} for a more detailed description of each of these stellar feedback modes.

\section{A New Module for Jets} \label{sec:jets_module}

\begin{figure}[t!]
\centering
\includegraphics[width = \linewidth]{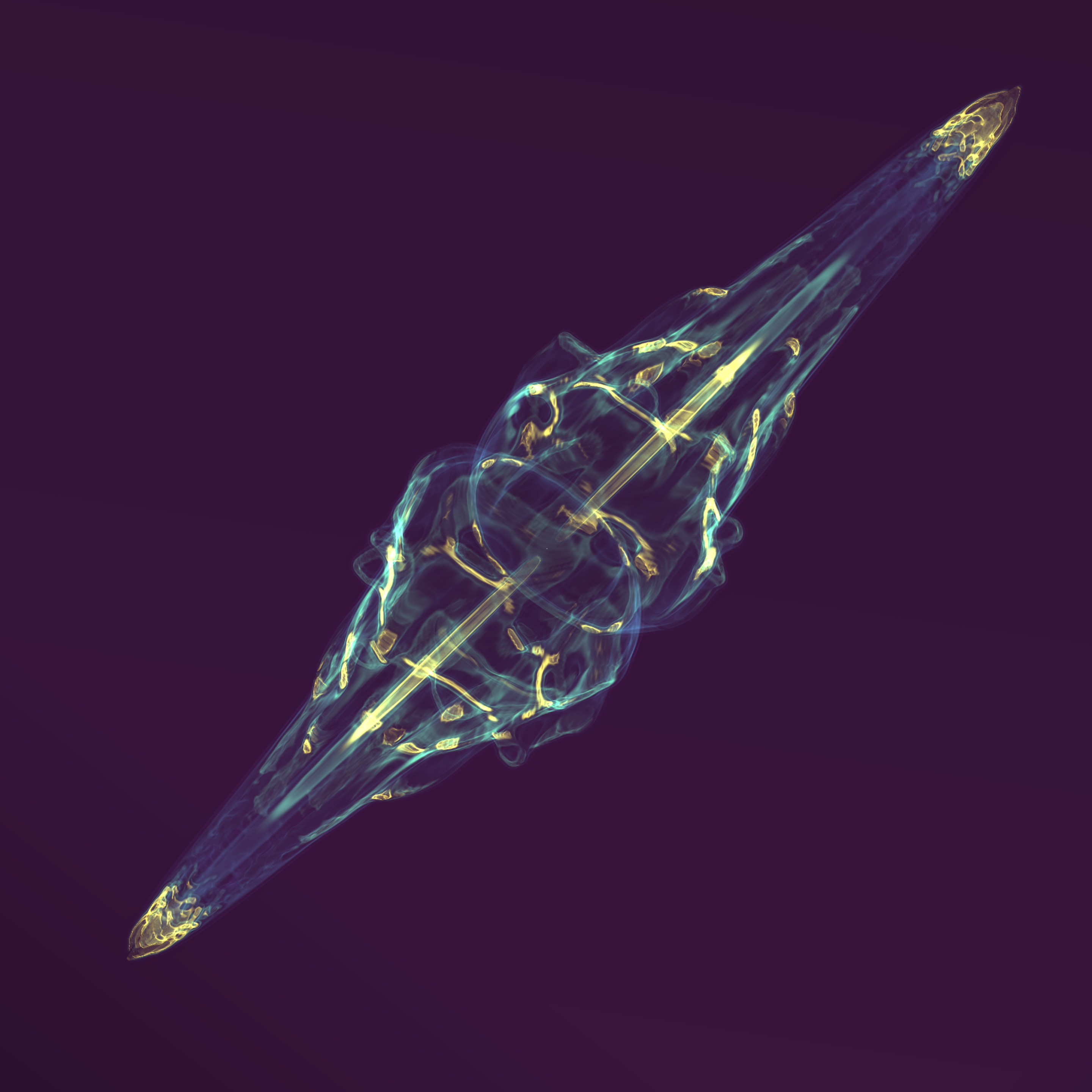}
\caption{A volume rendering of a protostellar jet produced by a single $5~\msun$ star particle in a uniform $2.18 \times 10^{-23}$~g~cm$^{-3}$ density background after 25.5~kyr. This volume rendering was generated from a simulation run using the jets branch of the \torch\ code (commit \href{https://bitbucket.org/torch-sf/torch/commits/d43eab7f0c3cc506d4e0b38b9d938d9ce993fb22}{d43eab7}). 
Note that although the injection weighting is smooth, as described in Section~\ref{sec:rad_ang_dep}, hydrodynamical instabilities develop as the injected material propagates into the denser background material.
\label{fig:jet_rendering}}
\end{figure}

The new protostellar jet module is an extension of the pre-existing stellar winds module.
The resulting wind and jets injection routine injects either a stellar wind or a protostellar jet based on the age and mass of the star particle.
In the protostellar jets case, the updated module introduces angular and radial dependence to the outflow (see Subsection~\ref{sec:rad_ang_dep}) and sets the mass injection rate $dm/dt$ and injection velocity $v_{\mathrm{jet}}$ for the protostellar outflow (see Subsection~\ref{sec:dmdt_vel}).
Our sub-grid model for protostellar jet injection is adapted from that of \citet{Cunningham+2011}.

Both the mass and the age of each star particle are used to distinguish between star particles that produce winds (generally, high mass stars), star particles that produce jets (generally, low mass stars), and star particles that produce neither (very low mass stars or low mass stars that are done injecting a jet).
Jets are injected for only the length of a short jet-injection phase at the beginning of a star particle's lifetime, as set by the user defined parameter \verb|jet_time| (see Subsection~\ref{sec:jet_params}).
Which star particles inject a wind is set by a user defined parameter that is separate from the jets parameters; any star particle with a mass greater than or equal to \verb|min_wind_mass| injects a wind.
For star particles that are producing winds, the injection method uses a spherically symmetric injection region \citep{Wall+2020} and injects stellar winds for the lifetime of the star after the completion of any jet phase.

A volume rendering of the density field for a protostellar jet injected by a single star particle in a uniform density box is shown in Figure~\ref{fig:jet_rendering}.
This run injected a protostellar jet from a 5.0~$\msun$ star for 25.5~kyr into a $2.18 \times 10^{-23}$~g~cm$^{-3}$ background medium (corresponding to a number density of approximately 10 cm$^{-3}$) with a minimum cell size of 0.0156~pc and a box size of 16~pc.
This run used the default jet parameters discussed below and listed in Table~\ref{tab:user_params}.
By the time of the snapshot shown in Figure~\ref{fig:jet_rendering}, the protostellar jet extends more than 3~pc away from the star particle in each direction.
This run demonstrates the evolution of a jet injected with our model when in isolation (i.e., no other feedback mechanisms or variations in gas density), although with much higher refinement than in a full cloud run.
At this resolution, the hydrodynamical instabilities that develop as the injected material propagates into the background material are easily visible in the evolution of the gas, although not all of these instabilities will be resolved in the full cloud simulations below.
We discuss the impact of the maximum refinement level on a single jet injected into a uniform medium in Appendix~\ref{sec:single_jet_resolution} and on the overall cloud evolution in Appendix~\ref{sec:jets_resolution}.
The shape of the injection region for the jet is discussed further in Subsection~\ref{sec:rad_ang_dep}.

This section is organized as follows: Subsection~\ref{sec:jet_params} describes the top-level, user-defined parameters that set the properties of the jets model; Subsection~\ref{sec:accretion} describes how the accretion process has been updated to account for the mass injected by jets; Subsection~\ref{sec:ang_mom} describes how the direction of the jet is determined, based on the angular momentum of the star particle; Subsection~\ref{sec:dmdt_vel} describes how the mass injection rate and the jet velocity are determined; and Subsection~\ref{sec:rad_ang_dep} describes the shape of the injection region for the jets model.

\subsection{Jet Parameters} \label{sec:jet_params}

\begin{table*}
    \centering
    \caption{Description of user defined parameters for the jets module.}
    \begin{tabular}{lcl}
        \hline \hline 
        Param. Name & Default Value & Description\\ \hline
         \verb|min_jet_mass| & 1.0 M$_{\odot}$ & minimum mass that produces a jet \\
         \verb|max_jet_mass| & 7.0 M$_{\odot}$ & maximum mass that produces a jet \\
         \verb|jet_time| & 100 kyr & length of time over which jet is injected \\
         \verb|jet_mass_fraction| & 0.33 & mass injected by jet as a fraction of the star's mass \\
         \verb|jet_vel_fraction| & 0.25 & jet velocity as a fraction of the Keplerian velocity \\
         \hline
    \end{tabular}
    \tablecomments{The default values for our jet module are given. If jets are not included in a model, the variables \texttt{jet\_mass\_fraction} and \texttt{jet\_time} should be set to 0 to ensure the mass injection rate vanishes.}
    \label{tab:user_params}
\end{table*}

Table~\ref{tab:user_params} lists the five user defined parameters for the jets module that can be accessed through the \verb|flash.par| file along with their default values in the publicly available version of \torch\ code, and which are used for the fiducial runs presented here.
These parameters can be used to fine-tune the protostellar jet model.
We briefly review these parameters here, and subsequent sections further detail how they are used.

The range of stellar masses that will inject a protostellar jet is determined by \verb|min_jet_mass| and \verb|max_jet_mass|. 
This mass range is independent of the minimum masses for injecting winds, radiative feedback, or supernovae.
Note that a given star particle cannot produce both jets and winds simultaneously.
If the mass ranges for jets and winds overlap, then jets are injected for the length of time set by \verb|jet_time|, after which the star particle injects a spherical wind.

Our default mass range for jet injection is 1--7~$\msun$.
This default range reflects two simplifying assumptions.
First, since winds from massive stars will inject a greater amount of momentum and energy than a protostellar jet from the same star, we choose for our fiducial runs to have wind-producing stars only inject winds. 
We set the minimum mass for stellar winds to the same value as the minimum mass for ionizing radiation, which is the minimum sufficient to form an \ion{H}{2} region.
This matches the typical minimum wind injection mass for \torch~runs of 7.0~$\msun$ \citep[see, for instance][]{Wall+2020, Cournoyer-Cloutier2021, Cournoyer-Cloutier2023, Cournoyer-Cloutier2024, Cournoyer-Cloutier2024b}.
Second, the number of stars goes up as mass decreases, while jet momentum decreases. 
Thus, we expect the computational expense of including jets from lower mass stars to grow with decreasing mass, while the momentum injected by those star will be small.
Therefore, for the fiducial runs presented here, we choose to limit jet injection to stars of at least 1~$\msun$.
The relative impact of jets from different mass ranges, including low mass stars and wind-producing stars should be investigated in future work.

The lifetime of each jet is determined by the \verb|jet_time| parameter. 
We have set the \verb|jet_time| to be constant for all stars and therefore independent of the mass of the star particle.
Since observations indicate that protostars have different timescales for their outflows \citep{Shepherd+1996}, this is a simplifying assumption of our model.
Our default value for the lifetime of the jet injection is 100~kyr, reflecting the typical main accretion phase of intermediate mass stars \citep{Frank2014, Bally2016}.
Since \torch~does not model the protostellar phase, we opt to use a constant jet lifetime and inject the jet during the first part of the star particle's main sequence lifetime.
Our use of a constant jet lifetime is an area for future investigation and is discussed further in Section~\ref{sec:discuss_model_params}.

The \verb|jet_mass_fraction| parameter determines the total amount of mass injected by a given protostellar jet and is defined as a fraction of the zero-age main sequence mass of the star particle.
For instance, if \verb|jet_mass_fraction| is set to $1/3$, then a 6$M_{\odot}$ star will have a protostellar jet that injects 2$M_{\odot}$ of material over its lifetime.
Observations suggest that the fraction of mass that is injected by the jet depends on the accretion rate and the mass of the protostar \citep{Bally2016, Nisini2018}.
This value is often reported as a ratio of the jet injection rate to the accretion rate.
Theory and observations find values for this ratio that range from $\sim$0.01 to 0.5 \citep[see, e.g.,][]{Beuther2002, Banerjee2006, Price2012, Ellerbroek2013, Federrath2014, Bally2016, Nisini2018}. 
For a constant jet injection time, this ratio can be used to estimate our mass fraction.
For the purpose of demonstrating our model, we use a large value that is still well within the proposed range.
The use of a constant mass fraction is an area that would benefit from future exploration, as discussed further in Section~\ref{sec:discuss_model_params}.

The \verb|jet_vel_fraction| parameter sets the velocity of the jet and is defined as a fraction of the Keplerian velocity at the surface of the star, based on its mass and radius.
The stellar properties used for this calculation are determined by the \verb|seba| code.
A value of \verb|jet_vel_fraction| = 1.0 would correspond to 100\% of the Keplerian velocity at the surface of the main-sequence star.
This is the maximum physical velocity for a protostellar jet. 

In fact, we expect the actual injection velocity for protostellar jets to be much lower since jets are most likely launched by a protostar, which will have a rather larger radius than a main-sequence star of the same mass, and from the inner edge of a protostellar disk around that protostar rather from the surface of the protostar \citep[e.g.,][]{Tomisaka2002}. 
Therefore, we halve the injection velocity once for the increased radius of a protostar relative to its corresponding main sequence star, and again for the distance between the surface of the protostar and the inner region of the disk, giving a default value of \verb|jet_vel_fraction| = 0.25. 
This is a rough value chosen to account for the understood physics of how jets are launched \citep[e.g.,][]{Tomisaka2002}.

Indeed, for a 7~$\msun$ star (the highest mass of our default jet mass range), the Keplerian velocity will be approximately 550~km~s$^{-1}$ \citep[where we have used the empirically derived mass-radius relation from][to get an approximate radius]{Eker+2018}.
Considering the observations discussed in Section~\ref{sec:intro-jets}, this suggests that while the main-sequence Keplerian velocity is not entirely unreasonable for the jet injection velocity, it is rather fast.

\subsection{Modifying the Sink Accretion}  \label{sec:accretion}

The overall star formation routine for \torch\ is described in Section~\ref{sec:star_formation}.
The jet module modifies the star formation routine in order to ensure overall mass conservation for the simulation.
The jet module requires the sink particle to accrete the total mass of the star \textit{and} the total mass of the jet prior to forming a new star. 
The jet mass is equal to $(\verb|jet_mass_fraction| \times M_{*} )$. 
And so, in order to form a star, the sink must accrete $( M_{*} \times (1+ \verb|jet_mass_fraction|)) $.

The star particle forms with only the initial stellar mass and the excess accreted mass is then injected by the jet over a period of time (set by \verb|jet_time|).
This results in a slight temporal disconnect between the accretion process and the jet injection process.
This also means that mass is temporarily not conserved as the mass of the jet is initially removed from the grid.  
However, over the course of the whole stellar lifetime, mass is still conserved.
Since the mass for the jet is accreted by the sink in addition to the required stellar mass, the sampling of the IMF is unchanged from previous versions of \torch.

Note that this approach of accreting the jet mass in addition to the stellar mass contrasts with the stellar winds routine in which the mass injected by the wind is due to mass loss from the star particle.
Star particles that produce jets, however, do not undergo mass loss due to the jets as the mass for the jets is accreted by the sink particle before forming the star particle.

\subsection{Angular Momentum of Star Particles} \label{sec:ang_mom}

The angular momentum of each sink particle is updated in accordance with the net angular momentum of its accreted material. 
While \torch\ did not previously implement an angular momentum property for star particles, our jet model links the orientation of the jet to the rotation axis of the star particle. 
To implement this, we add an angular momentum orientation for star particles, which is inherited from the angular momentum of the sink that formed the star particle and used to set the direction of the jet.

A large portion of the angular momentum of the gas that forms a star is dissipated during the star formation process, so the angular momentum of the star does not conserve the angular momentum of the original gas \citep[see e.g.,][and references therein]{Tomisaka2002, Misugi2024}.
Fully modeling this process is beyond the scope of the current star formation implementation.
Thus, the angular momentum vector for the star particle is set to a unit vector based on the angular momentum direction of the sink when the star forms and the angular momentum of the star remains constant after formation.

There are additional, sub-grid processes that alter stellar rotation axes during the formation process, but are not accounted for in our model, such as tidal circularization or mass transfer in a binary \citep[see][for a recent review]{Marchant2024} or continued turbulent accretion \citep{Fielding2015, Offner2016, Kong+2019}.
We account for these processes by perturbing the direction of each star's angular momentum vector on formation by multiplying each component by a random number that is drawn from a Gaussian distribution with mean 1.0 and standard deviation 0.1, prior to normalizing the vector to magnitude 1.
By using a Gaussian distribution, we retain a preference for jets to be aligned with the angular momentum of the accreted material, but allow for jets to occasionally be misaligned.
This random perturbation also prevents many stars being formed on the same timestep with perfectly parallel outflows.
Although some alignment of protostellar outflows has been observed \citep[][]{Stephens+2017,Kong+2019}, perfect alignment would be unphysical and, indeed, \citet{MakinFroebrich2018} find distributions consistent with random orientations.

The angular momentum magnitude of the star particle is a unit value and has no physical meaning in our setup, but the sink particle angular momentum magnitude does reflect the cumulative angular momentum of the accreted gas.
Thus, we reduce the sink particle angular momentum to account for the loss of angular momentum by star formation. 
The magnitude of the sink particle angular momentum is reduced in proportion to the fraction of the sink particle's initial mass that is transferred to the star particle and its corresponding jet:
\begin{equation}
    |L_{\mathrm{final}}| = |L_{\mathrm{init}}| \left( \frac{m_{\mathrm{sink}} - (m_{\mathrm{star}} +m_{\mathrm{jet}})}{m_{\mathrm{sink}}} \right)  \ .
    \label{eq:ang_momentum_update}
\end{equation}
The sink's angular momentum direction is not modified when a star particle forms since the star particle has the same orientation (excluding the random perturbation).
This approach ensures that the sink particle angular momentum is reduced to zero if its mass is ever reduced to zero and that the angular momentum magnitude does not grow arbitrarily large as gas is accreted.

Our method orients the protostellar jets according to the angular momentum of the stars' progenitor gas.  
We note that the transfer of angular momentum magnitude from gas to sink to stars is not fully self-consistent and could be improved in future work.

\subsection{Jet Mass Injection and Velocity} \label{sec:dmdt_vel}

Two quantities control the properties of the injected material: the mass injection rate $dm/dt$ and the jet injection velocity $v_{\mathrm{jet}}$.
The mass injection rate is set by the total mass injected by the jet and the jet lifetime:
\begin{equation}
    \frac{dm}{dt} = \frac{M_{*} \times \mathrm{\texttt{jet\_mass\_fraction}}}{ \mathrm{\texttt{jet\_time}}} \ \ .
    \label{eq:dmdt}
\end{equation}

The velocity of the jet is set to be a fraction of the Keplerian velocity $v_{\rm K}$ at the surface of the star:
\begin{align}
    v_{\mathrm{jet}} & = \mathrm{\texttt{jet\_vel\_fraction}} \times v_{\mathrm{K}} \ \ ,
    \label{eq:jet_vel}
\end{align}
where
\begin{align}
     v_{\mathrm{K}} &= (G M_{*}/R_{*})^{1/2} \\
    &= 550\,\mbox{km s}^{-1}\times
    \left(\frac{M_{\star}}{7\,{\rm M}_{\odot}}\right)^{1/2}\,
    \left(\frac{R_{\star}}{4\,{\rm R}_{\odot}}\right)^{-1/2} 
    \label{eq:kep_vel}
\end{align}
\citep[using the mass-radius relation in][]{Eker+2018}.
In this way, the injection velocity of the jet varies with the mass of the star.

\subsection{Radial and Angular Dependence} \label{sec:rad_ang_dep}

\begin{figure}[b!t]
\centering
\includegraphics[width = \linewidth]{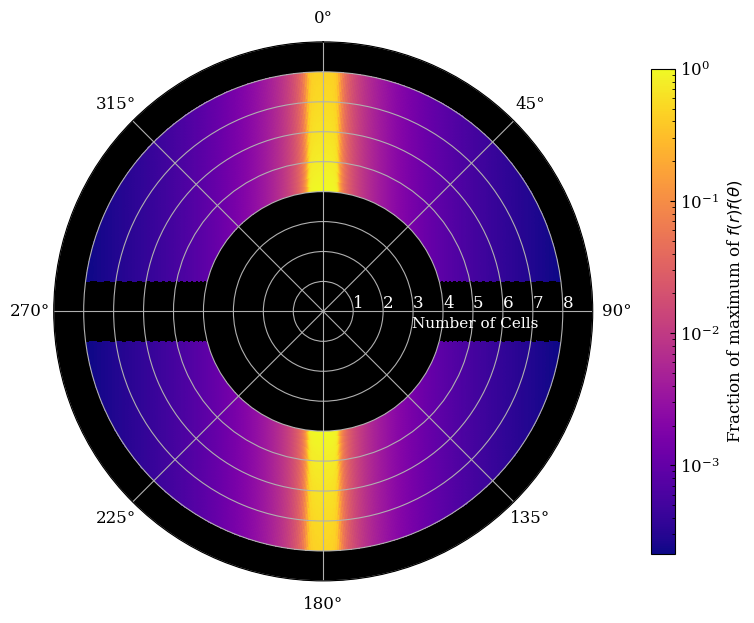}
\caption{A plot of the injection region described in Section~\ref{sec:rad_ang_dep}. The heatmap shows the relative value (as a fraction of the maximum value) of the product of the radial and angular dependence: $f(r)g(\theta)$. The actual weighting for each cell of the injection region is the product of this quantity evaluated at the cell center and the geometric factor determining the overlap of the injection region and the specific cell, and is normalized such that all the weights add to 1. Neither this normalization nor the geometric factor are included in this plot, in order to focus on the distribution of the $f(r)g(\theta)$ factor. Regions shaded black are outside the injection region and have a weighting of 0.
\label{fig:inj_region}}
\end{figure}

Our subgrid model for the shape of the injection region follows the model described in \cite{Cunningham+2011}, with some adaptations.
Here, we describe the key features of our adapted model.

For each cell within the injection region, the radial $r$ and angular $\theta$ positions of the cell with respect to the angular momentum axis of the star particle are used to determine the fraction of the jet energy and momentum that is injected in that cell.
Thus, the jet is aligned with the angular momentum direction of the star particle (as described in Subsection~\ref{sec:ang_mom}).
The injection region is symmetric with rotation about the angular momentum vector.

The weighting for how much material is injected in each cell $i$ in the injection region is determined by:
\begin{equation}
    w_i = f(r_i)g(\theta_i) F_i \ ,
    \label{eq:weighting}
\end{equation}
where $F_i \in [0,1]$ is the fraction of cell volume that overlaps with the injection region and the radial and angular dependencies are given by $f(r_i)$ and $g(\theta_i)$.
After calculating $w_i$ for each cell in the injection region, we normalize the weights to sum to unity over the entire injection region.
The total mass injected and the overall injection velocity are determined by the mass injection rate $dm/dt$ and the jet injection velocity $v_{\mathrm{jet}}$ described in the previous subsection.
In this way, the portion of the mass injected in each cell is determined by the cell weighting and the normalization step ensures that the correct total amount of mass is injected.
Note that our normalization method differs from that of \cite{Cunningham+2011}, who use the constants $c_1$ and $c_2$ to normalize the injection strength; we keep these constants in the following equations for comparison but set the values to 1.0 as they are unneeded with our normalization method.

Material is injected onto the grid within a small region close to the star particle, but with a gap right next to the star particle.
This is set by the function for the radial dependence:
\begin{equation}
    f(r) = 
\begin{cases}
 \frac{1}{c_1 r^{2} } \ \  & 4 \Delta_{\mathrm{min}} \leq r \leq 8 \Delta_{\mathrm{min}}\\
 0  & \mathrm{otherwise} \ ,
\end{cases}
\label{eq:fr}
\end{equation}
where $\Delta_{\mathrm{min}}$ is the minimum cell size (we require the injection region to be maximally refined).
This is equivalent to Eq.~19 in \cite{Cunningham+2011}.

\begin{widetext}

The function that describes the angular dependence $g(\theta)$ also creates a gap near the equator.
This gap approximates the presence of an accretion disk that would prevent the injection of jet material in that direction.
The angular dependence takes the form:
\begin{equation}
    g(\theta) = 
\begin{cases}
 \frac{1}{c_2 \, \Delta\theta} \left( \frac{1}{\theta_{0} \, \sqrt{1+\theta_{0}^{2} }} \right) \times \psi (\theta)
 & |\sin(\pi/2 - \theta)| \geq \Delta_{\mathrm{min}} / r \\
 0  & \mathrm{otherwise} \ 
\end{cases}
\label{eq:ftheta}
\end{equation}
\citep[based on Equations\ 21 and 22 of][]{Cunningham+2011}, where $\psi(\theta)$ is given by:
\begin{equation}
    \psi (\theta)= 
 \left[ 
 \tan^{-1} \left( \frac{\sqrt{1+\theta_0^{2}} \, \tan(\theta + \Delta\theta/2) )} {\theta_0} \right) - \right. 
 \left. \tan^{-1} \left( \frac{\sqrt{1+\theta_0^{2}} \, \tan(\theta - \Delta\theta/2) )} {\theta_0} \right)
  \right]  \ \ 
  \label{eq:psi}
\end{equation}
\citep[based on Equation\ 22 of][]{Cunningham+2011}. In our implementation, we have used $\Delta\theta = \Delta_{\mathrm{min}} / r$. 
Note that this value of $\Delta \theta$ differs from that implemented in \cite{Cunningham+2011}, who use a constant $\Delta\theta = \arctan(1/8)$. 
We find that this difference has no significant effect on the resultant jet; we discuss this difference in our implementation and its implications in Appendix~\ref{sec:dtheta}.
Future runs with \torch\ that include jets should use an updated value of $\Delta \theta = \arctan(1/8)$ to align with the \cite{Cunningham+2011} model.
\end{widetext}

The parameter $\theta_{0}$ sets the strength of the angular dependence of the jet injection region, with larger values corresponding to a weaker angular dependence and thus to a larger effective opening angle. 
For instance, $\theta_0 = 10$ produces an injection region with essentially no angular dependence, whereas $\theta_0 = 0.01$ results in strong collimation (as in Figure~\ref{fig:inj_region}).
We use $ \theta_{0} = 0.01$, following \cite{Matzner1999} and \citet{Cunningham+2011}.
We opt to use the same value of $\theta_0$ for all of our runs, although varying the opening angle of the jet is an interesting area for future exploration.

\section{Exploring Jet Parameter Variation} \label{sec:testing}

\begin{deluxetable*}{lcccccccr}[thb!]
\tabletypesize{\footnotesize}
\tablecaption{Summary of the runs presented in this paper. The simulation suite includes runs with and without protostellar jets, with different jet parameters, and with two different cloud masses. 
The single star simulation used for Figure~\ref{fig:jet_rendering} is not listed here due to its substantially different setup.}
\label{tab:testruns}
\tablecolumns{5}
\tablewidth{0pt}
\tablehead{
\colhead{Sim.\ Name} & 
\colhead{$M_{\mathrm{cloud}}$ ($\msun$)} & 
\colhead{$\Sigma_{\mathrm{init}}$ ($\msun \mathrm{pc}^{-2}$)} & 
\colhead{Min.\ Cell (pc)} & 
\colhead{Jets} & 
\colhead{Jet Time (kyr)} & 
\colhead{Jet Vel.} &
\colhead{Final SFE (\%)} & 
\colhead{$N_{\mathrm{*}}$}
}
\startdata
    \texttt{S-nojet1}  & $5 \times 10^{3}$  & $32.5$  & $0.137$      & Off & ---   & ---  & $32.3 $ & $2,975$ \\
    \texttt{S-nojet2}$^{\dagger}$   & $5 \times 10^{3}$  & $32.5$  & $0.137$      & Off & ---  & ---  & $12.6 ^{\dagger}$ & $1,077^{\dagger}$ \\
    \texttt{S-jet1} & $5 \times 10^{3}$  & $32.5$  & $0.137$      & On  & $100$ & 0.25  & $14.3 $ & $1,150$ \\
    \texttt{S-jet2} & $5 \times 10^{3}$  & $32.5$  & $0.137$      & On  & $100$ & 0.25  & $8.29 $ & $547$ \\
    \texttt{S-jet3} & $5 \times 10^{3}$  & $32.5$  & $0.137$      & On  & $100$ & 0.25  & $12.6 $ & $1,110$ \\
    \texttt{S-short}   & $5 \times 10^{3}$  & $32.5$  & $0.137$      & On  & $50 $ & 0.25  & $21.7 $ & $1,690$ \\
    \texttt{S-slow}   & $5 \times 10^{3}$  & $32.5$  & $0.137$      & On  & $100$ & 0.125 & $27.7 $ & $2,304$ \\
    \texttt{S-fast}   & $5 \times 10^{3}$  & $32.5$  & $0.137$      & On  & $100$ & 0.5   & $12.3 $ & $985$ \\
    \texttt{M-nojet}   & $2 \times 10^{4}$  & $130 $  & $0.137$ & Off & ---   & --- & $47.9 $ & $16,716$ \\
    \texttt{M-jet} & $2 \times 10^{4}$  & $130 $  & $0.137$ & On  & $100$ & 0.25  & $32.1 $ & $10,299$ \\
    \texttt{M-short}   & $2 \times 10^{4}$  & $130 $  & $0.137$ & On  & $50$  & 0.25  & $31.7 $ & $9,893$ 
\enddata
\tablecomments{\textit{Sim.\ Name}: name of simulation used in the text;  $M_{\mathrm{cloud}}$: initial total mass of the spherical cloud; $\Sigma_{\mathrm{init}}$: initial surface density at the center of the sphere; \textit{Min.\ Cell}: minimum cell size; \textit{Jets}: whether protostellar jets are turned on; \textit{Jet Time}: value of \texttt{jet\_time}; \textit{Jet Vel.}: value of \texttt{jet\_vel\_fraction}; \textit{SFE}: integrated star formation efficiency after 3~\tff; $N_{\mathrm{*}}$: the total number of stars formed after 3~\tff. $^{\dagger}$The SFE and $N_{\mathrm{*}}$ for the \snojettwo~run are measured at 5.9~Myr.}
\end{deluxetable*}

\begin{deluxetable}{ll}[b!]
\tabletypesize{\footnotesize}
\tablecaption{Constant parameters for the simulation suite.
\label{tab:sims_const}}
\tablecolumns{5}
\tablewidth{0pt}
\tablehead{
\colhead{Parameter} &
\colhead{Value}
}
\startdata
    Cloud Radius ($R$) & 7~pc \\
    Box size ($L$)  & 17.5~pc \\
    $\alpha_{\mathrm{vir}}$ & 0.5 \\
    B-field ($B_z$) & 3~$\mu$G \\
    \texttt{min\_jet\_mass} & 1.0~$\msun$  \\
    \texttt{max\_jet\_mass} & 7.0~$\msun$  \\
    \texttt{jet\_mass\_fraction} & 0.33  \\
    \texttt{min\_wind\_mass} & 7.0~$\msun$ \\
\enddata
\tablecomments{Cloud radius, B-field, and $\alpha_{\mathrm{vir}}$ denote the initial values for the spherical cloud. Box size denotes the full edge length of the cubical simulation domain. These parameters are the same for all runs in Table~\ref{tab:testruns}.
}
\end{deluxetable}

We demonstrate our new jets module, and the importance of including protostellar jets in simulations of star formation, using a suite of 13 \torch~simulations with varying initial cloud masses and protostellar jet parameters.
In this section, we describe the setup of the simulation suite, and in Sections~\ref{sec:results}~and~\ref{sec:discussion}, we present our analysis of the simulations and of the impact of protostellar jets.
The data files for each snapshot and key set up files for each run are available in a Globus Collection that can be found, along with movies of density and temperature projection plots, at doi:\href{https://doi.org/10.5531/sd.astro.10}{10.5531/sd.astro.10} \citep[][]{Appel25data}.

Table~\ref{tab:testruns} lists the variable parameters, and Table~\ref{tab:sims_const} lists the constant parameters, for each of our runs.
We present runs for spherical clouds of two different masses: $5 \times 10^{3}~\msun$ (the small-cloud runs, labeled with \texttt{S}) and $2 \times 10^4~\msun$ (the medium-cloud runs, labeled with \texttt{M}).
We refer to our higher-mass clouds as medium clouds since the mass of the \texttt{M} clouds in the present work corresponds to the mass of the lowest-mass (M1) clouds in \cite{Cournoyer-Cloutier2024b}.
We kept the radius of the cloud (7~pc) and the box size ($L/2 = 8.75$~pc) consistent for all the runs; thus, the two cloud masses correspond to two different density regimes.
All of the runs use the same random seed for the initial turbulent velocity field of the gas and for the star mass lists used for the star formation routine (although differing numbers and accretion rates of sinks will still yield different star formation histories).
We choose our random seeds and other initial conditions to be consistent with the initial conditions used in previous work with \torch\ \citep[in particular,][]{Cournoyer-Cloutier2024b}.

All of the runs were initialized with a virial parameter of $\alpha_{\mathrm{vir}}=0.5$.
The virial parameter represents the ratio of the kinetic energy $T$ and gravitational potential energy $U$ of the cloud, and for a spherical cloud is
\begin{equation}
    \alpha_{\mathrm{vir}} = \frac{2 T }{|U|} = \frac{5 \sigma ^2 R}{G M} \ ,
\end{equation}
where $R$ is the radius of the cloud, $M$ is the mass of the cloud, and $\sigma$ is the velocity dispersion of the cloud \citep{Bertoldi1992,Cournoyer-Cloutier2023}.

We set up each run with a uniform magnetic field of 3~$\mu$G in the $z$-direction.
Following the approach in \cite{Crutcher1999}, this results in a mass-to-flux ratio at the center of the sphere of:
\begin{equation}
    M_{\mathrm{sm}}/\Phi = \frac{32.5~\msun \mathrm{pc}^{-2}}{3~\mu G} = 2262 ~\frac{g}{G \ cm^{2}}
\end{equation}
for the smaller clouds and 
\begin{equation}
    M_{\mathrm{med}}/\Phi = \frac{130~\msun \mathrm{pc}^{-2}}{3~\mu G} = 9050 ~\frac{g}{G \ cm^{2}}
\end{equation}
for the higher mass clouds, where we have used the approximate initial surface density at the center of the sphere and the initial magnetic field strength.
\cite{Mouschovias1976} derive an expression for the critical mass for a cloud to collapse, from which we can determine the critical value of the mass-to-flux ratio \citep[see also][]{Crutcher1999, Beuther2018, Polak2024a}:
\begin{equation}
    \left(M/\Phi\right)_{\mathrm{crit}} = \frac{0.53}{3 \pi} \left(\frac{5}{G}\right)^{1/2} = 487 ~\frac{g}{G \ cm^{2}}
\end{equation}
This means that both clouds are significantly super-critical at the start of the runs, by factors of approximately 4.6 and 18.6.
Thus, we expect very little support from magnetic fields for these runs.

Each run (except \snojettwo) was allowed to progress for 3 free-fall times (\tff) and Table~\ref{tab:testruns} reports the SFE and $N_{\mathrm{*}}$ values at the end of the runs. 
For the \texttt{S} runs (with $5 \times 10^{3}~\msun$) \tff$=2.2$~Myr.
For the \texttt{M} runs (with $2 \times 10^4~\msun$) \tff$=1.1$~Myr.
\snojettwo\ was ended early (at $\sim5.9$~Myr) due to increasing numerical instabilities at later times.

For each cloud mass, we present a run without jets (\snojet, \mnojet), a fiducial run with jets using the set of default jet parameters defined in Table~\ref{tab:user_params} (\sdef, \mdef), and a run with a halved jet lifetime (\sshort, \mshort).
For the lower density cloud, we also include runs with slower (\sslow) and faster (\sfast) jet injection velocities in order to explore the space of reasonable values for this velocity.
Future work should explore justifications for, and the implications of, different values for each of the user defined parameters.

In addition to runs which vary the jet parameters, we also present additional runs with identical parameters (\snojettwo, \sdeftwo, \sdefthree) to explore the stochastic variability of individual runs.
\snojettwo\ is set up to be identical to \snojet.
\sdeftwo\ and \sdefthree\ are set up to be identical to \sdef.
This includes consistent random seeds between all of our runs for both the initial turbulent velocity field and the star mass lists that are generated for the sink particles.
However, other sources of randomness persist in our runs, including due to the ray-tracing module \texttt{FERVENT} and due to numerical noise.
These sources of randomness can result in very different star formation histories between runs with identical initial conditions.
We compare the runs with identical initial conditions in Section~\ref{sec:small} and discuss the impact of randomness on our runs further in Section~\ref{sec:random}.

In addition to the runs presented in Table~\ref{tab:testruns}, we ran two additional simulations with lower (\slo) and higher (\shi) maximum refinement values, which are presented and discussed in Appendix~\ref{sec:jets_resolution}.

\section{The Impact of Jets on Star Formation} \label{sec:results}

\begin{figure*}[p!]
\centering
\includegraphics[width =0.95\linewidth]{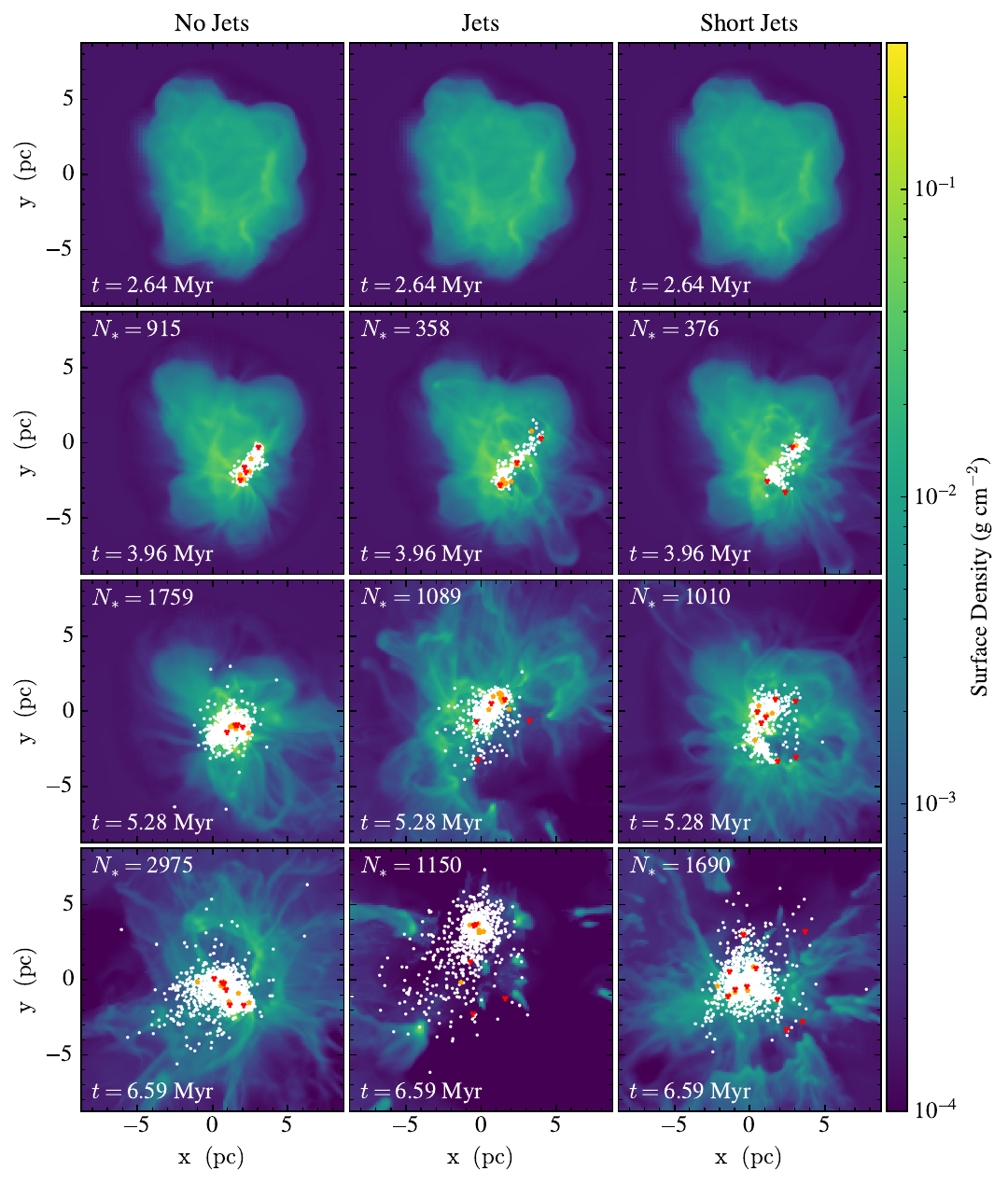}
\caption{Projection plots of density along the z-axis are shown for the \snojet, \sdef, and \sshort~runs at four points in time. 
The projected x-y position of each star particle is shown as a white dot for non-feedback stars and an orange star for any star that is currently injecting either winds or jets. Sink particles are shown as red Y symbols. The time for each plot and the number of star particles is shown on each panel. For these runs $t_{\mathrm{ff}} = 2.2$~Myr, so the plotted times correspond to approximately 1.2~\tff, 1.8~\tff, 2.4~\tff, and 3.0~\tff\ since the start of the run. Movies of the projected density for each of the runs presented in this paper can be found, along with the Globus Collection for the data in this paper, at doi:\href{https://doi.org/10.5531/sd.astro.10}{10.5531/sd.astro.10} \citep[][]{Appel25data}.
\label{fig:s_projections}}
\end{figure*}

\begin{figure}[t!]
\centering
\includegraphics[width = \linewidth]{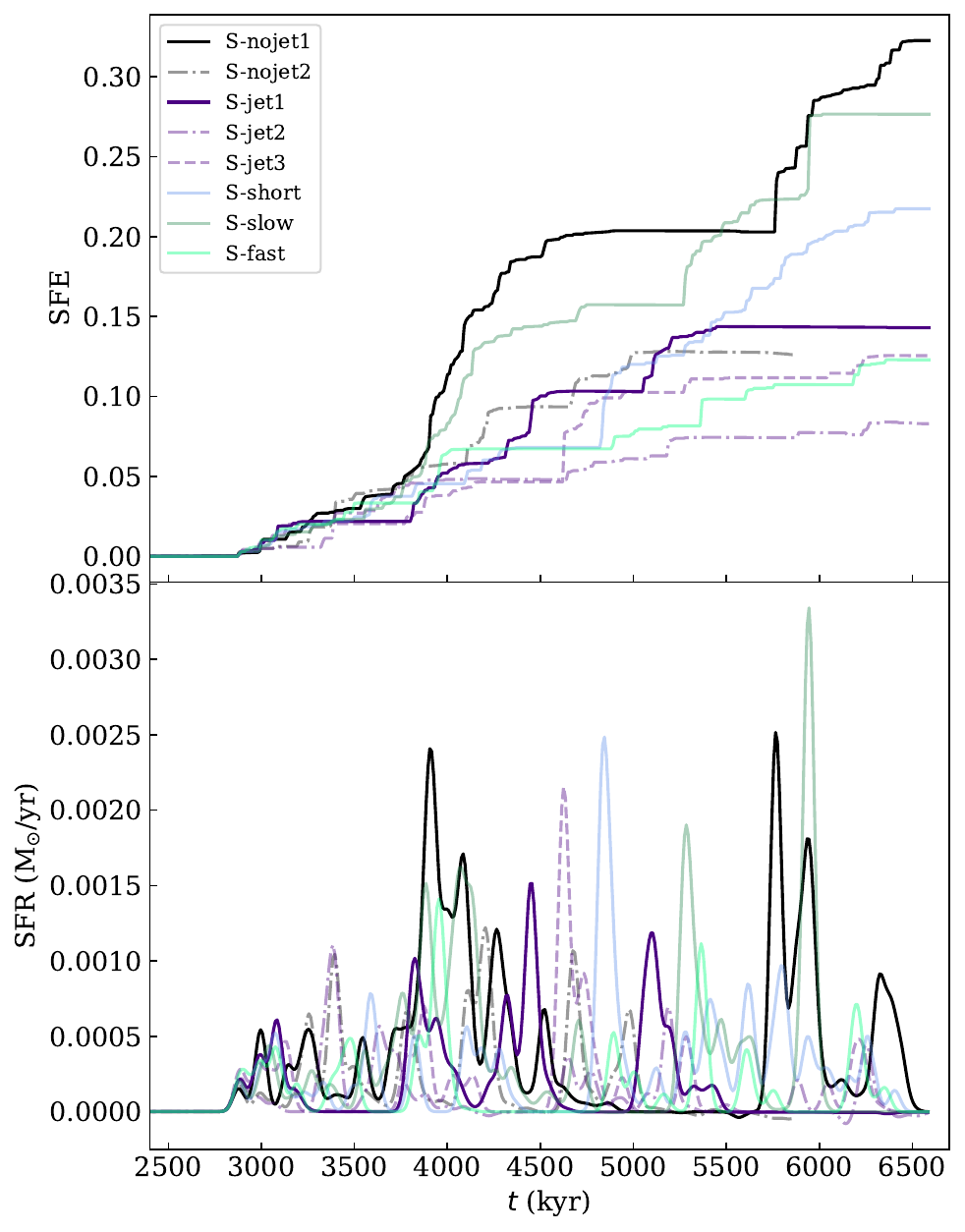}
\caption{A comparison of the integrated SFE and the smoothed SFR as functions of time for the eight runs with smaller initial clouds (and consistent maximum refinement) described in Section~\ref{sec:testing}. The runs without jets are shown in black (\snojet\ as a solid line and \snojettwo\ as a dotted line) while the six runs with different jets parameters are shown in lines of various colors. 
The lower panel shows the SFR smoothed with a Gaussian filter with a standard deviation of 3, which corresponds to approximately 30~kyr.
\label{fig:sm_sfe}}
\end{figure}

\begin{figure}[t!]
\centering
\includegraphics[width = 0.95\linewidth]{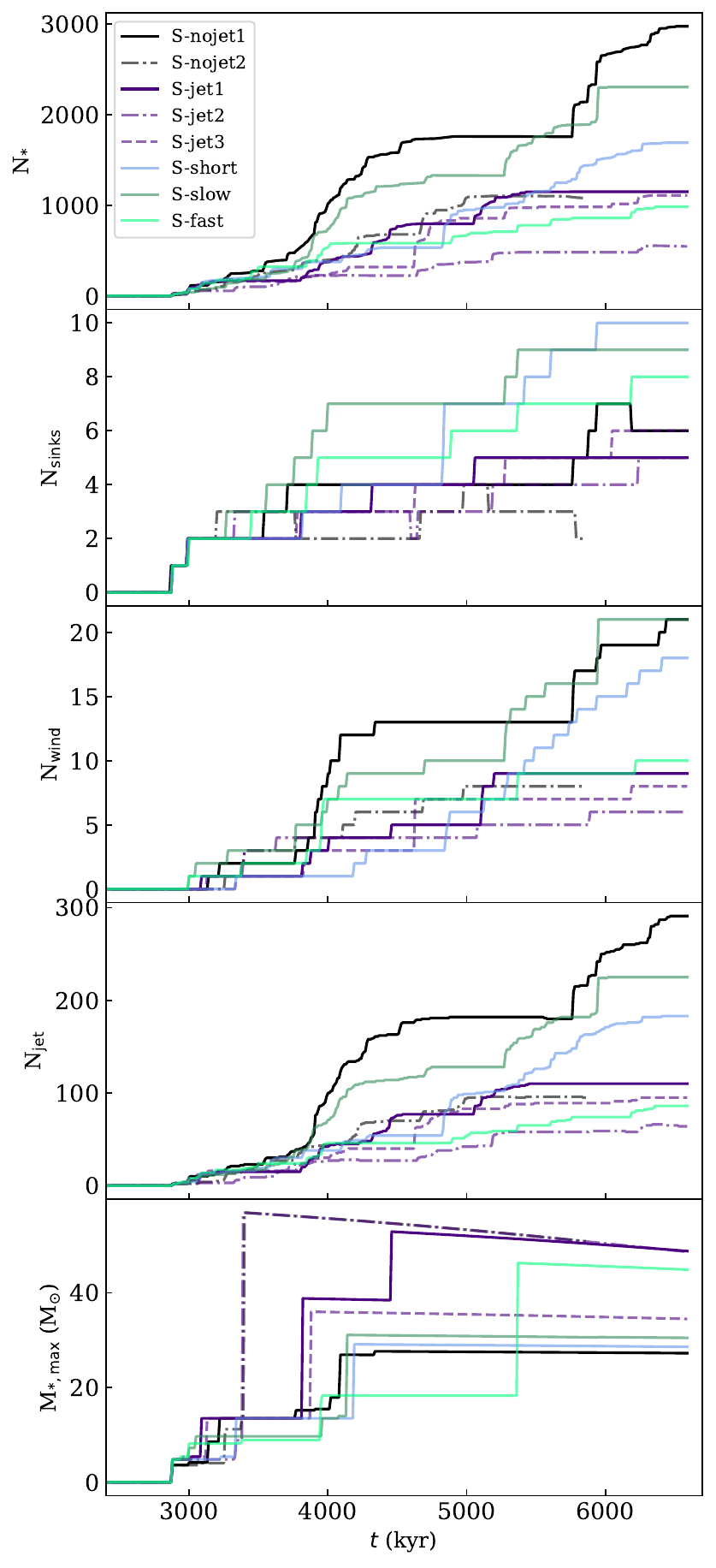}
\caption{ A comparison of several star formation diagnostics as functions of time for the same runs shown in Figure~\ref{fig:sm_sfe}. From top to bottom the panels show: the total number of star particles, the total number of sink particles, the number of wind mass stars ($M \geq 7~\msun$), the number of jet mass stars ($1~\msun \leq M < 7~\msun$), and the most massive star on the grid (in $\msun$). Two of the runs (\snojettwo\ and \sdeftwo) form a very massive wind star at around $3400$~kyr, which significantly impacts the subsequent evolution of both runs.
\label{fig:sm_stars}}
\end{figure}

\subsection{The Small Clouds} \label{sec:small}

Figure~\ref{fig:s_projections} shows projection plots of the density field for three of the runs with lower mass clouds (\snojet, \sdef, \sshort) at four different times.
The runs start from the same initial conditions and so progress identically at early times. 
The first stars that form are in similar places. 
However, the runs diverge significantly at later times.
In the bottom three rows, the material ejected by the jets is visible in the diffuse gas for the \sdef~and \sshort~runs, especially in the lower right corner of the plots.
From these plots, we can see that jets act to disrupt the gas cloud starting soon after the onset of star formation.

Figure~\ref{fig:sm_sfe} shows the integrated SFE and the smoothed SFR as functions of time for these runs as well as the other lower mass runs. 
The SFR is smoothed using a Gaussian filter with a standard deviation of 3 data points, corresponding to approximately 30 kyr.
As with the projection plots in Figure~\ref{fig:s_projections}, the runs start similarly before diverging at later times. 
\snojet\ exhibits the fastest growth of the SFE as well as the highest final SFE.
\snojet\ also has several large spikes in the SFR, although other runs also exhibit large spikes the SFR; this is a consequence of sink particles depositing several stars in a single timestep at various points in the runs.
The various runs with jets (\sdef, \sdeftwo, \sdefthree, \sshort, \sslow, and \sfast) vary in how rapidly the SFE evolves, with final SFE values ranging from 8.29\% to 27.7\%.
The runs with jets do not undergo a faster growth in the SFE than \snojet, except for the \sshort~run, which briefly has a higher SFE than \snojet around 5.5~Myr before the \snojet\ run undergoes a burst of star formation.

Figure~\ref{fig:sm_stars} shows several other star formation diagnostics over time for these same runs, including the total number of star particles, the total number of sink particles, the number of stars that will produce winds, the number of stars that could produce a jet (if jets are on and regardless of whether the star is currently producing a jet at a given time), and the most massive star on the grid.

Each of the runs in Figures~\ref{fig:sm_sfe}~and~\ref{fig:sm_stars} follows a similar evolutionary progression: they form stars relatively slowly initially before gaining stellar mass more rapidly, have a highly variable SFR, and gain more sink particles as the run progresses.
At early times (before about 4000 kyr), the SFE and the total number of stars is fairly similar between all of the runs. 
At later times, however, the \snojet~run develops a higher total number of stars (and SFE) than all of the runs with jets.
Although the runs with different jet parameters do vary in terms of the evolution of the stellar population as well as the SFE, the notably higher SFE for the \snojet\ run without jets suggests that the presence of jets generally suppresses star formation.

In addition, the number of sink particles produced in the \snojet\ run generally falls in the lower end of the range of numbers of sinks produced by the various \texttt{S} runs, despite forming the most stars (see Figure~\ref{fig:sm_stars}). 
Although the number of sink particles formed in a particular run is somewhat stochastic and is sensitive to the evolution of the gas, this suggests that the runs with jets undergo more fragmentation, thus forming more separate regions where sink particles can form.
This agrees with results from \citet{Guszejnov2021} and \citet{Appel+2023}, both of whom find indications that protostellar jets increase gas fragmentation.

The \sshort\ run tests the impact of halving the lifetime of the jet (set by \verb|jet_time|). 
Since the other jet parameters are kept the same, this also results in doubling the $dm/dt$, while keeping the jet injection velocity the same. 
Figures~\ref{fig:sm_sfe}~and~\ref{fig:sm_stars} suggest that the shorter jets are similarly effective at reducing star formation at early times as the default, but that they become less effective at later times.
Prior to around 5.5~Myr, the \sshort\ run progresses very similarly to the \sdef\ run.
However, after this point, the SFE of the \sshort\ run outpaces the \sdef\ run and produces a larger number of stars.
The comparatively high SFE and number of stars for the \sshort\ run at late times indicates that injecting jets for a longer time increases the duration over which jet feedback is effective at reducing star formation and highlights the importance of future work in understanding jet lifetimes in order to correctly track star cluster formation at later times.

The \sfast\ and \sslow\ runs explore the impact of increasing and decreasing the jet injection velocity while keeping the jet lifetime and jet mass the same.  
Figures~\ref{fig:sm_sfe}~and~\ref{fig:sm_stars} suggest that the jet velocity has a significant impact on how effectively jets can inhibit star formation.
Once the runs start to diverge, the \sfast\ run generally has the lowest SFE and number of stars (excepting the \sdeftwo\ run, which is discussed below).
In contrast, the \sslow\ run has the highest SFE and number of stars of the various runs with jets and is the most similar to the the \snojet\ run, even briefly showing a higher SFE near 5.5~Myr.
This indicates that altering the jet injection velocity significantly alters how effective jet feedback is at slowing the progression of star formation.
This is likely due to the increased (decreased) momentum of the injected material with an increased (decreased) injection velocity, which results in more (less) momentum and energy injected into the gas.

The second run without jets, \snojettwo, has a much lower SFE and SFR, even at later times, than \snojet~and several of the runs with jets. 
This star is more massive than any of the stars formed by most of the runs, especially at such early times.
The formation of high mass stars in our runs (such as the high mass star in \snojettwo) is a consequence of how we handle the star formation prescription, as described in Section~\ref{sec:star_formation}.
We further discuss the implications of how this prescription is impacted by sources of randomness in our runs in Section~\ref{sec:random}.

The formation of a high mass star early in the \snojettwo~run is especially important in light of work by \cite{Lewis+2023}, who show that early forming massive stars have a dramatic impact on subsequent star formation.
Indeed, the way that star formation nearly halts in \snojettwo~shortly after the formation of a $\sim57~\msun$ star agrees with these results.

Figure~\ref{fig:sm_stars} shows that \sdeftwo~also forms a high mass star at almost exactly the same time that \snojettwo~forms a high mass star.
Given that the formation of a high mass star significantly alters the progression of star formation, the similarity between these two runs provides an excellent test to separate the impact of jets and early forming massive stars.
In particular, we can directly test the role of jets in the case where there is an early forming massive star by comparing \sdeftwo~and \snojettwo.
And we can contrast this with  the role of jets in \sdef~and \snojet, where a massive star does not form at early times.
The reduction in star formation in the \sdef\ run relative to \snojet\ is discussed above.
Comparing \sdeftwo\ and \snojettwo\ in Figure~\ref{fig:sm_sfe} (the two dotted lines) shows that the inclusion of protostellar jets slightly reduces the growth of the SFE in these runs as well.
Thus, although the impact of the wind from a high mass star greatly overpowers the impact of protostellar jets (e.g., comparing \sdef~and \snojettwo), the inclusion of protostellar jets does still reduce star formation if factors such as the formation of a high mass star are constant (e.g., comparing \sdeftwo~and \snojettwo).

\subsection{The Medium Clouds}

\begin{figure*}[p!]
\centering
\includegraphics[width =0.95\linewidth]{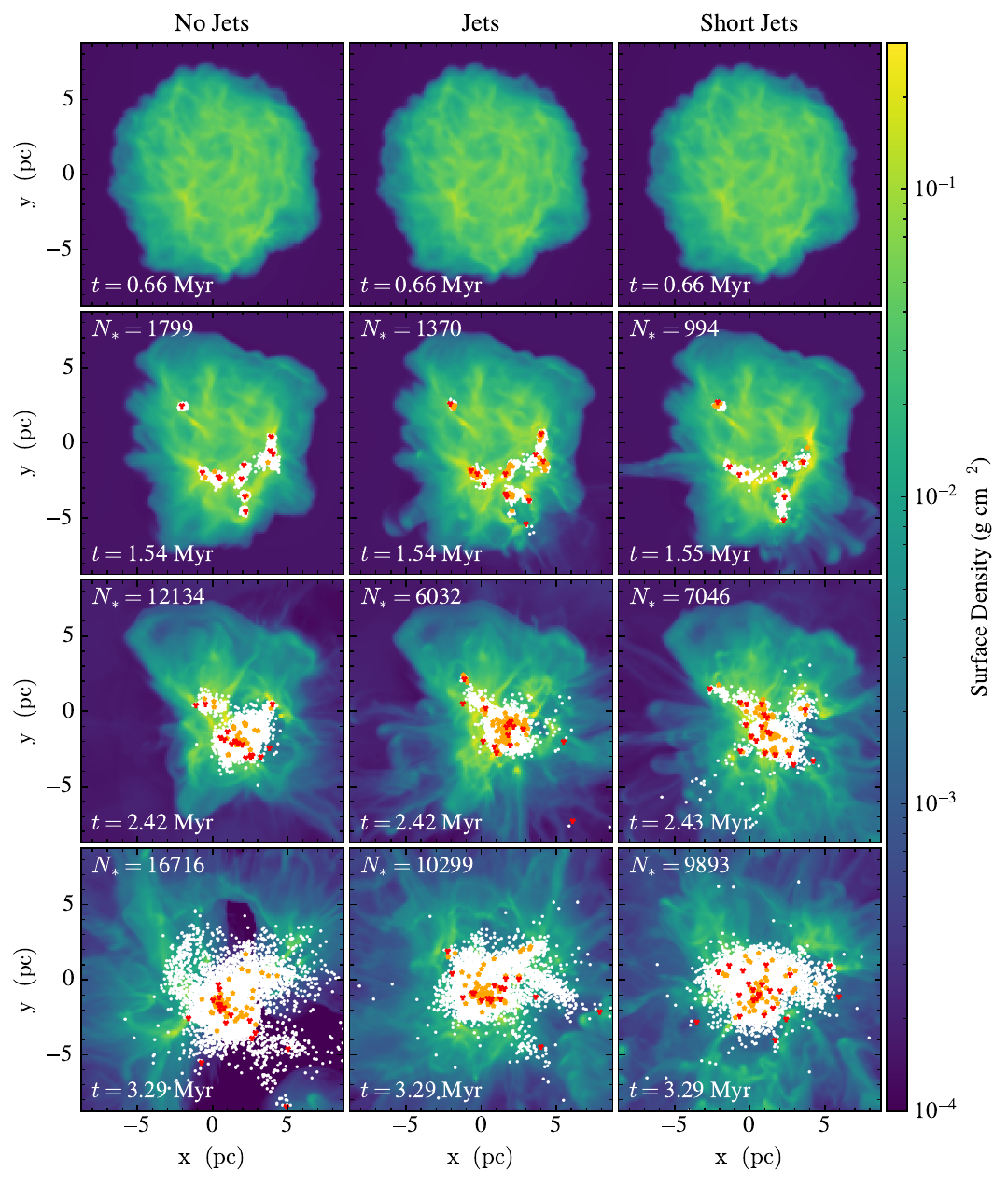} 
\caption{Projection plots of density along the z-axis are shown for the \mnojet, \mdef, and \mshort~runs at four points in time. 
The projected x-y position of each star particle is shown as a white dot for non-feedback stars and an orange star for any star that is currently injecting either winds or jets. Sink particles are shown as red Y symbols. The time for each plot and the number of star particles is shown on each panel. For these runs $t_{\mathrm{ff}} = 1.1$~Myr, so the plotted times correspond to approximately 0.6~\tff, 1.4~\tff, 2.2~\tff, and 3.0~\tff\ since the start of the run. Movies of the projected density for each of the runs presented in this paper can be found, along with the Globus Collection for the data in this paper, at doi:\href{https://doi.org/10.5531/sd.astro.10}{10.5531/sd.astro.10} \citep[][]{Appel25data}.
\label{fig:m_projections}}
\end{figure*}

\begin{figure}[t!]
\centering
\includegraphics[width = \linewidth]{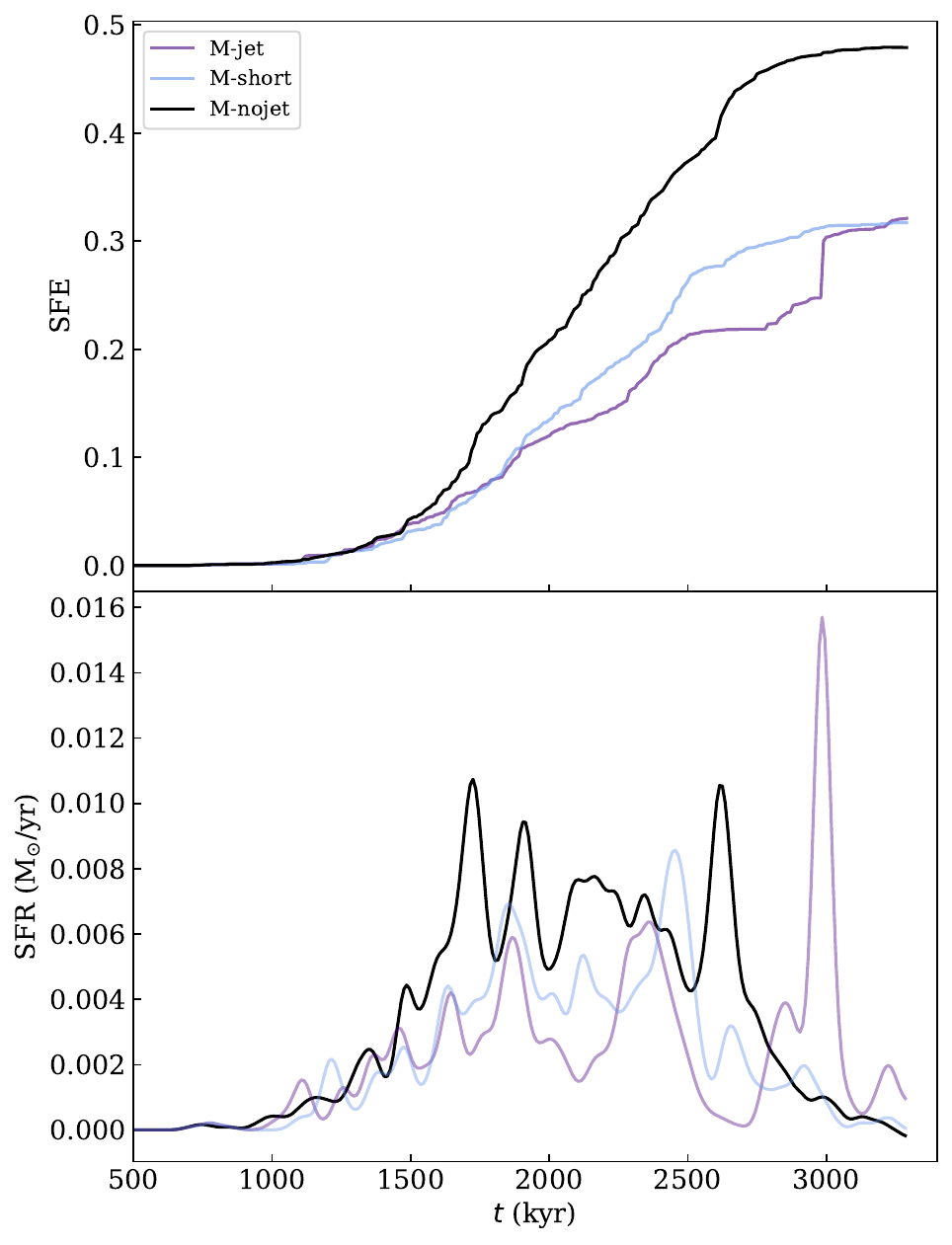}
\caption{ A comparison of the integrated SFE and the smoothed SFR as functions of time for the three runs with a medium initial cloud mass, as described in Section~\ref{sec:testing}. The run without jets (\texttt{M-nojet}) is shown in black while the two runs with different jets parameters are shown in lines of various colors.  The lower panel shows the SFR smoothed with a Gaussian filter with a standard deviation of 3, which corresponds to approximately 30~kyr.
Although the three runs are similar at early times, the runs with protostellar jets form stars more slowly than the run without jets at later times.
\label{fig:med_sfe}}
\end{figure}

\begin{figure}[t!]
\centering
\includegraphics[width = 0.95\linewidth]{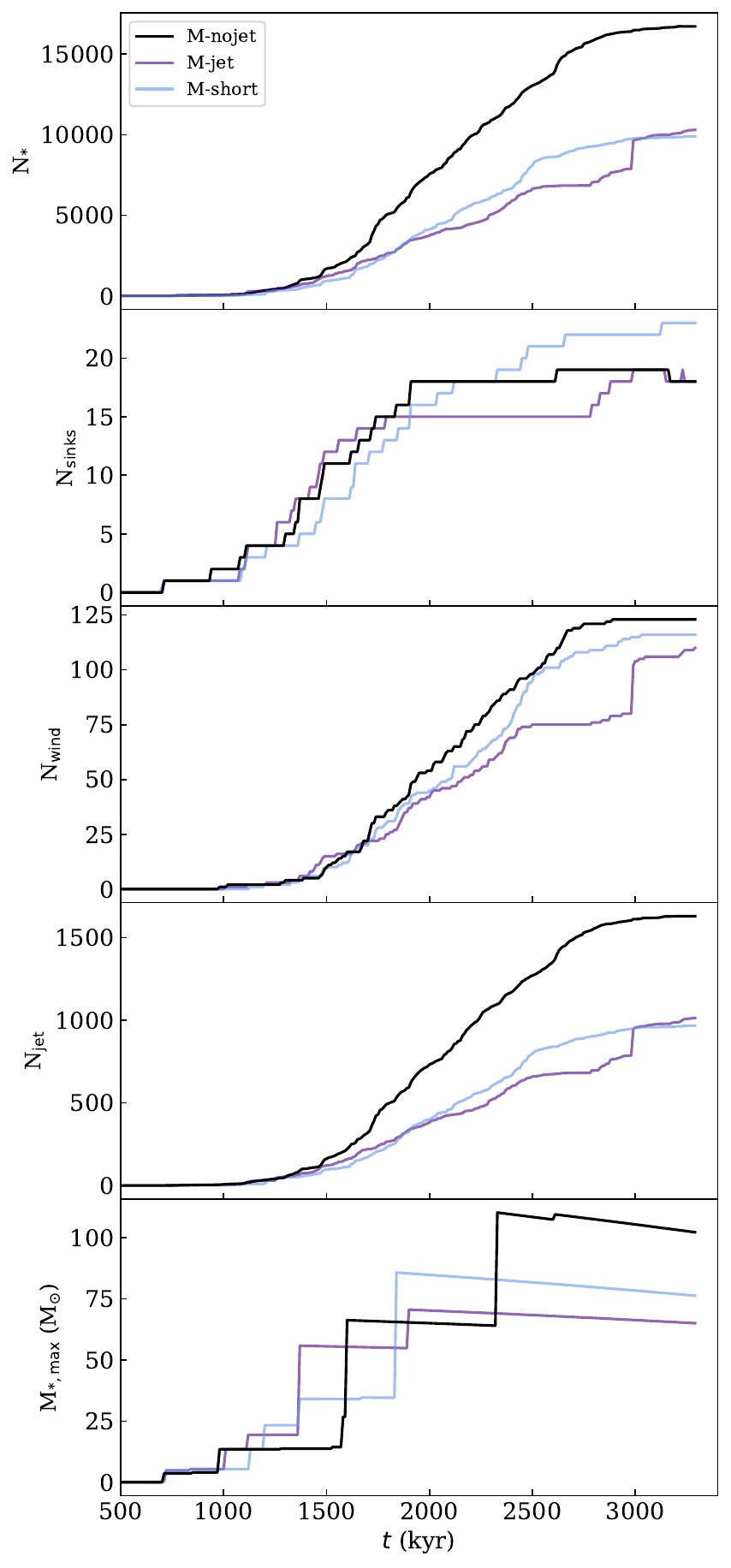}
\caption{A comparison of several star formation diagnostics as functions of time for the same runs shown in Figure~\ref{fig:med_sfe}. From top to bottom the panels show: the total number of star particles, the total number of sink particles, the number of wind mass stars ($M \geq 7~\msun$), the number of jet mass stars ($1~\msun \leq M < 7~\msun$), and the most massive star on the grid (in $\msun$). 
\label{fig:med_stars}}
\end{figure}

Figure~\ref{fig:m_projections} shows projection plots of the density field for the three runs with higher density clouds (\mnojet, \mdef, and \mshort).
The runs start identically to each other and start forming stars much sooner than the \verb|S| runs due to the higher starting densities.
However, the runs diverge significantly at later times.
The runs with and without jets differ at later times by as much as a factor of two in the total number of stars formed: at 3~\tff\ the \mdef~run has 61\% of the number of stars in the \mnojet\ run and only 50\% as many stars at 2.2~\tff.
The middle two rows of Figure~\ref{fig:m_projections} also demonstrate noticeable differences in where stars form.
The material ejected out of the cloud by the jets is visible in the middle two rows for the \mdef~and \mshort~runs.
Future work should further investigate the influence of protostellar jets on the properties and distribution of star clusters and subclusters.

Figure~\ref{fig:med_sfe} shows the integrated SFE and the smoothed SFR as functions of time for these runs (similar to Figure~\ref{fig:sm_sfe}) and Figure~\ref{fig:med_stars} shows several star formation diagnostics over time (similar to Figure~\ref{fig:sm_stars}).
All three runs begin by forming stars slowly, before developing a more rapid and variable SFR.
All three runs rapidly gain sink particles, reaching as much as twice as many sink particles as the \verb|S| runs.
However, around 1.5~Myr, when the SFR begins to grow rapidly, the two runs with protostellar jets diverge noticeably from the run without protostellar jets, forming far fewer stars.
Indeed, in these higher density clouds, we find that the inclusion of protostellar jets significantly slows down star formation.
This can be seen in the consistently lower integrated SFE and generally lower SFR at later times when jets are included.
However, the two runs that include protostellar jets are overall quite similar to each other.
The \mshort\ run develops a slightly higher SFE than the \mdef\ run between 2~myr and 3~Myr, but the SFE of the \mdef\ run catches up at late times. 
Overall, this suggests that star formation in these clouds is not especially sensitive to the chosen lifetime of the jets.
Future work should investigate the role of protostellar jets at later times and in higher density, more massive clouds.

\subsection{Energetics with and without Jets} \label{sec:energy}

\begin{figure*}[t!]
\centering
\includegraphics[width = 0.9\linewidth]{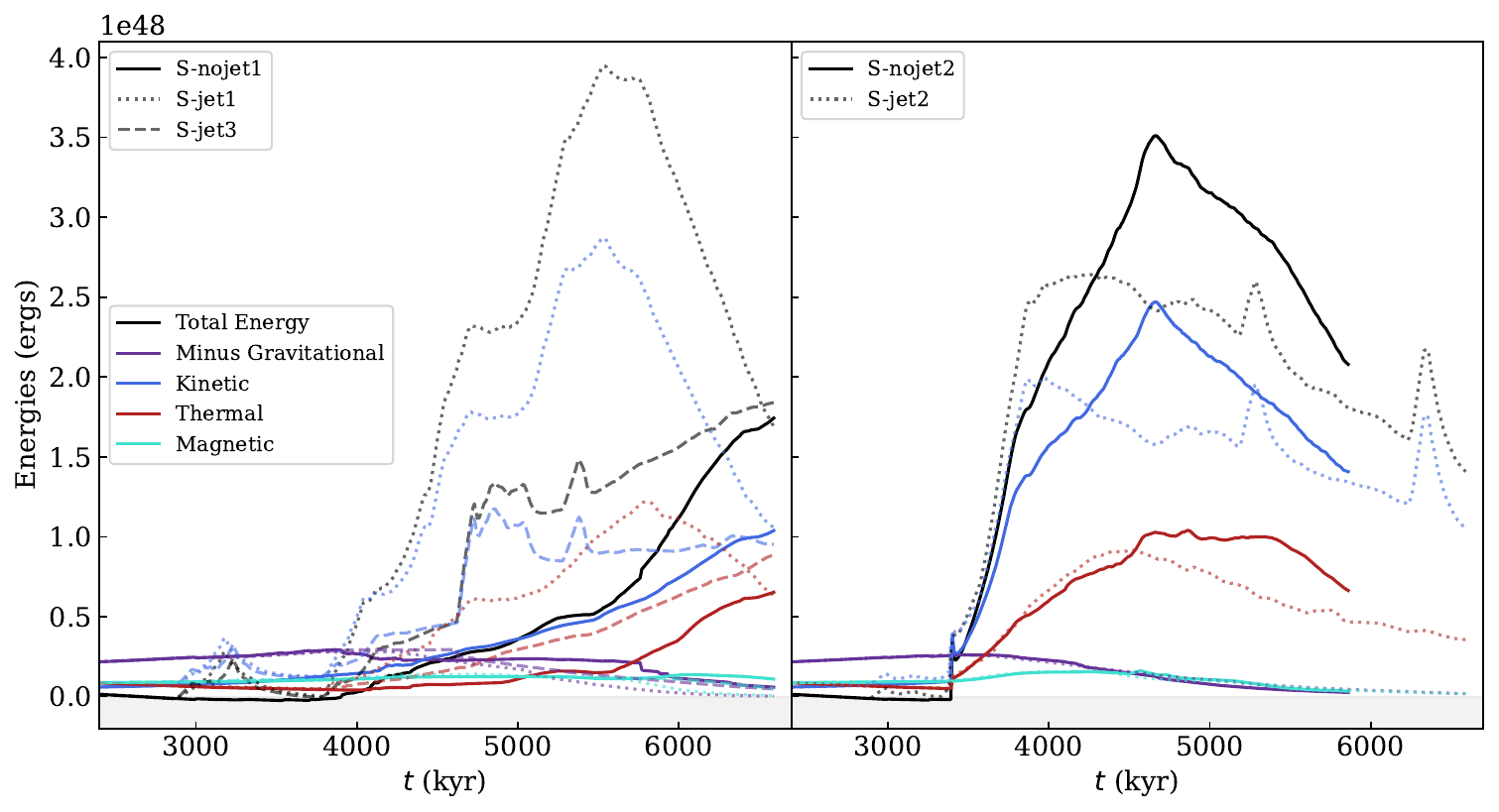}
\caption{ A comparison of thermal energy ({\em red}; Equation~\ref{eq:thermal}), kinetic energy ({\em blue}; Equation~\ref{eq:kinetic}), gravitational potential energy ({\em purple}; Equation~\ref{eq:grav}), magnetic energy ({\em turquoise}; Equation~\ref{eq:magnetic}), and total energy ({\em black}; Equation~\ref{eq:total}) for several of the small clouds. Note that the purple line plots the negative of the gravitational energy while the total energy accounts for the sign the gravitational energy.
The left panel shows the \snojet~run with solid lines and two of the runs with default jet parameters (\sdef~and \sdefthree) with dashed and dotted lines. 
The right panel shows the two runs that form a massive wind star early in the run, one with (\sdeftwo; dotted line) and one without (\snojettwo; solid line) jets. The massive wind stars significantly alter the subsequent progression of each run and so we consider these runs separately. 
In both panels, negative values implying gravitational boundedness are shaded gray.
\label{fig:sm_energy}
}
\end{figure*}

\begin{figure}[t!]
\centering
\includegraphics[width = \linewidth]{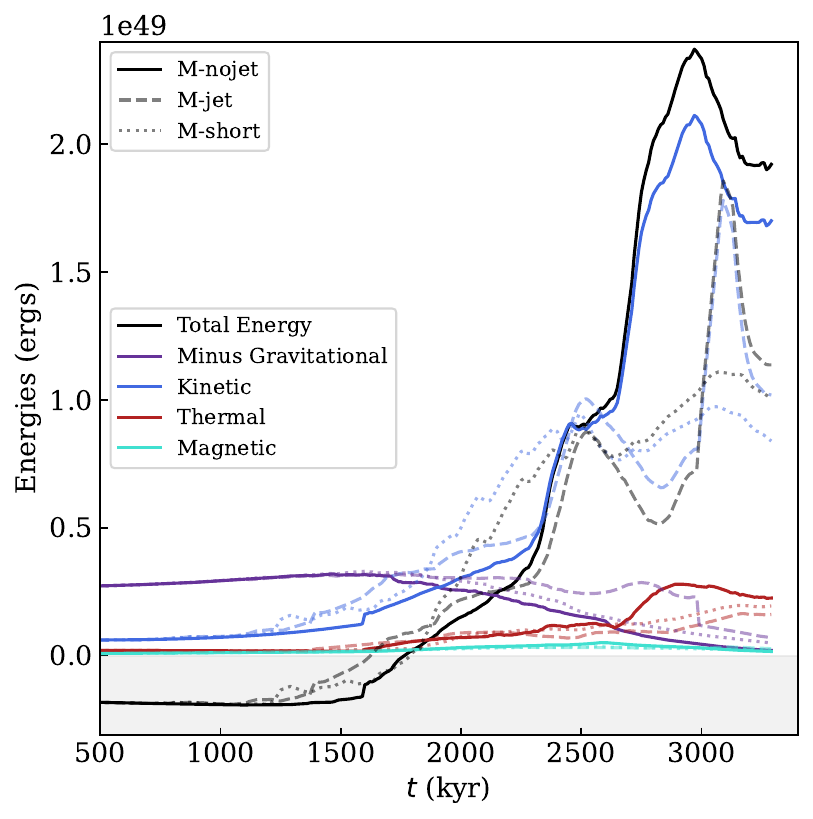}
\caption{ A comparison of different forms of energy, as in Figure~\ref{fig:sm_energy}, for the three medium clouds. The run without jets (\mnojet) is shown with solid lines, while the two runs with jets (\mdef~and \mshort) are shown with dashed and dotted lines. Note that the purple line plots the negative of the gravitational energy while the total energy accounts for the sign of the gravitational energy.  Negative values implying gravitational boundedness are shaded gray.
\label{fig:med_energy}}
\end{figure}

We explore the evolution of the kinetic energy (KE), thermal energy (TE), gravitational potential energy (GE), and magnetic energy (ME) of the gas in simulations with and without jets. 
We find that the inclusion of protostellar jets alters the evolution of the energy balance in the cloud.

We calculate the kinetic energy by summing the kinetic energy of the gas in all of the cells on the grid: 
\begin{equation}
    KE = \sum_{i} \frac{1}{2} m_i |v_i|^2 
    \label{eq:kinetic}
\end{equation}
where $m_i$ and $|v_i|$ are the mass and the magnitude of the velocity for cell $i$.
Similarly, we calculate the total thermal energy using:
\begin{equation}
    TE = \sum_i \frac{1}{(\gamma - 1)} \frac{m_i}{\mu \  m_p} k_b T_i
    \label{eq:thermal}
\end{equation}
where $T_i$ is the temperature of cell $i$, $m_p$ is the atomic mass of hydrogen, $k_b$ is the Boltzmann constant, and we assume $\gamma = 5/3$ and a mean molecular weight of $\mu=2.33$.
We calculate the gravitational potential energy analogously:
\begin{equation}
    GE =- \frac{1}{2} \sum_i \Phi_i m_i
    \label{eq:grav}
\end{equation}
where $\Phi_i$ is the gravitational potential in cell $i$.
Figures~\ref{fig:sm_energy}~and~\ref{fig:med_energy} plot the negative of the gravitational potential energy in order to more directly compare GE to the other sources of energy. 
Finally, we consider the total magnetic energy:
\begin{equation}
    ME = \sum_i  \mu_{i} V_i
    \label{eq:magnetic}
\end{equation}
where $\mu_{i}$ is the magnetic energy density in cell $i$, and $V_i$ is the volume for cell $i$, needed to convert from an energy density to a total energy.
The total energy in Figures~\ref{fig:sm_energy}~and~\ref{fig:med_energy} is calculated as:
\begin{equation}
    E_{\mathrm{tot}} = KE + TE + ME - GE \ \ .
    \label{eq:total}
\end{equation}
Thus, negative $E_{\mathrm{tot}}$ values correspond to the gas being gravitationally bound.

Figure~\ref{fig:sm_energy} shows the energy components for several of the \verb|S| runs with and without jets. 
Due to the dramatically different evolution of the runs that form an early massive star, the \snojettwo~and \sdeftwo~runs are shown separately (in the right hand panel) from the runs that did not form a massive star early on (left hand panel), so as to compare similar runs to each other.
For both sets of runs, the runs that included protostellar jets show an increase in the kinetic energy over the course of the simulation relative to the corresponding run without protostellar jets.
For the runs without an early-forming massive star (the left panel), the runs with jets show a dramatic increase in the kinetic energy relative to the run without jets.
The gravitational energy for \sdef\ shows a slight drop relative to the run without jets and the thermal energy shows a slight increase.
This suggests that the feedback from protostellar jets is both heating up and unbinding the cloud.

The reduced impact of protostellar jets on the energetics of the cloud in the case where the run forms an early massive star agrees with our expectation that an early-forming massive star will rapidly disrupt its host cloud \citep[e.g.][]{Lewis+2023} and the fact that winds from massive stars inject far more momentum and energy than protostellar jets from lower mass stars.

These same energy components are shown for the \verb|M| runs in Figure~\ref{fig:med_energy}. 
The thermal and magnetic energies are again relatively similar between runs with and without jets.
The kinetic energy, however, shows a significant increase with the inclusion of protostellar jets as the runs progress.
This results in a small increase in the total energy for the runs with protostellar jets.
In addition, for these higher mass clouds, the gravitational energy decreases faster for the run without protostellar jets.  

The energy analysis for each set of runs shows that protostellar jets stir the gas, increasing its kinetic energy \citep[see also][]{Appel+2022}.
This increase in the kinetic energy of the gas, in addition to removing gas mass from dense regions, suggests a mechanism by which the protostellar jets can reduce the SFR and slow the growth of the SFE, as seen in Figures~\ref{fig:sm_sfe}~and~\ref{fig:med_sfe}.

For both masses of clouds, we do not see a significant difference in the evolution of the magnetic energy with and without protostellar jets, so the jets are not driving substantial evolution of the magnetic energy through a dynamo or sweeping up field.
However, we do note that the magnetic energy is quite low compared to the gravitational energy, especially for our more massive clouds. 
This is likely a consequence of a relatively weak initial magnetic field (3~$\mu$G for all runs; see the discussion of the mass-to-flux ratio in Section~\ref{sec:testing}).
Indeed, the results from \cite{Crutcher1999} and \cite{Crutcher2010} show that our magnetic fields are unusually low for molecular clouds of these densities, indicating an avenue for future exploration.

\section{Discussion} \label{sec:discussion}

Overall, we find that the inclusion of protostellar jets impacts the evolution of star-forming regions, in agreement with previous work \citep[see, e.g.,][]{Federrath2015, Appel+2022, Grudic2022, Verliat2022, Appel+2023, Lebreuilly2024}.
We find that the inclusion of protostellar jets slows and suppresses star formation in both our lower- and higher-mass clouds.
Furthermore, the inclusion of protostellar jets alters the energetics of the clouds by increasing the total kinetic energy of the gas.
This aligns with \cite{Lebreuilly2024}, who find that jets increase the kinetic energy at small scales, and with \cite{Grudic2022}, who find that protostellar jets dominate the feedback momentum budget at early times.
Indeed, Figure~\ref{fig:sm_energy} suggests that the increase in kinetic energy due to protostellar jets begins to dissipate as the runs progress, meaning that the greatest impact of jets on the energetics is at early times.
We hypothesize that this increase in the kinetic energy is the main mechanism by which jets reduce the SFR and slow the growth of the stellar mass since this increase in the kinetic energy of the gas will oppose further collapse of the gas.

\subsection{Model Design and Parameters} \label{sec:discuss_model_params}

We explore a range of parameters for our protostellar jet model in our lower-mass clouds and find that altering the jet model parameters results in significant scatter in the behavior of our runs.
This highlights the importance of further exploration of the impact of each of these parameters on the formation and evolution of stellar clusters.
The default values of the user defined parameters are chosen to align with observed values, while the ability to vary these parameters ensures that our module can be used to test the impact of each of these parameters.
Indeed, the inclusion of protostellar jet feedback in \torch\ opens up the ability to address many new questions about feedback in star cluster formation and evolution. 

\subsubsection{Mass Range for Jets}

Our jets module is designed to inject jets for stars within some mass range, which is set by \verb|min_jet_mass| $< M <$ \verb|max_jet_mass|.
However, since stars of any mass can inject protostellar jets \citep[e.g.,][]{Shepherd+1996, Kolligan2018}, these parameters could be set so as to include jets from stars of any mass.
Our default value for \verb|max_jet_mass| is set to $7~\msun$ in order to account for the expectation that other modes of feedback are more impactful than jets for higher mass stars.
Our default value for \verb|min_jet_mass| is set to $1~\msun$ because including jets from lower mass stars increases the computational cost of each run due to the large number of low mass stars.
However, exploring the relative impact of protostellar jets from different mass protostars is an important area for future investigation.

\subsubsection{Jet Lifetimes}
As mentioned in Section~\ref{sec:intro-jets}, the length of time over which a protostellar jet is injected should be determined by the pre-main sequence evolutionary track of the star. 
In particular, the protostellar jet should be injected only for a short time at the beginning of the life of the star.
However, since pre-main sequence tracks are not yet implemented in \torch, we opt to use a fixed lifetime for all jets in our module.
Our \sshort\ run explores the impact of a shorter jet lifetime and finds that jets that are injected over a shorter period of time become less effective at slowing star formation at later times.
This demonstrates that the jet lifetime plays an important role in determining how jets impact the progression of star formation.

\subsubsection{Jet Direction}  

There is observational evidence (see Section~\ref{sec:intro}) to suggest that protostellar jets can change direction \citep[e.g.,][]{Shepherd+2000,Sai2024}.
Observations also show asymmetric jets, including unipolar outflows or multiple outflows \citep[e.g.,][]{Takaishi2024, Sai2024, Omura2024}.
Future development of the protostellar jet module could allow the jet direction to evolve in time based on subgrid models for disk-jet or binary star interactions.
A detailed comparison of our simulated jet orientations to observations of outflow orientations \citep[such as in][]{Stephens+2017,MakinFroebrich2018, Kong+2019} would also be an important future step in understanding how jet orientations are determined and how angular momentum is transferred from the ISM to stars.

We set the direction of the jet to be parallel to the angular momentum vector of the star particle, which is inherited from the gas accreted by the sink particle from which it forms (as discussed in Section~\ref{sec:ang_mom}).
While the sink particle angular momentum evolves as mass is accreted or lost through star formation, the angular momentum of each star particle, and thus the direction of each jet, is fixed.
We do not attempt to fully follow angular momentum conservation since this is unnecessary for implementing our protostellar jet model.
Thus, future enhancements of our jets module could be made to track the dissipation of angular momentum, including transferring angular momentum to the injected material of the jet.
Developing a more self-consistent treatment of the evolution of the angular momentum of the star particles, and thus the jet direction, would require treating each star as a sink particle, with accompanying difficulties in following accretion at disk and inner envelope scales where non-ideal MHD and non-ionizing radiative heating become important.

This would be of interest because there is controversy in the literature \citep[see e.g.,][]{Kong+2019} about whether protostellar jets and outflows are or are not perpendicular to the gas filaments within which the protostars form.
The variation may be a result of different gas environments or a result of stellar evolution processes.
Updating the prescription for the jet orientation could provide an opportunity to explore how different prescriptions determine the final protostellar jet orientation.

Furthermore, the current implementation of the jets module does not include binaries, although \cite{Cournoyer-Cloutier2021} has previously implemented primordial binaries in \torch, and dynamical binaries can also form in our runs \citep{Wall+2019}.
Observations suggest that binaries play a significant role in setting the direction and properties of protostellar jets \citep[e.g.,][]{Shepherd+2000, Fourkas2024, Sai2024}.
This should be explored in future work.

\subsubsection{Pre-main Sequence Tracks}

Another approximation that we use is to neglect pre-main sequence stellar evolution in \torch. 
Since our star particles are formed as zero-age main sequence stars, jets must be tied either to the sink particles (which accrete material but have no other properties of a star) or to the first part of the lifetime of a main sequence star.
We choose to link protostellar jets to the beginning of the main sequence track in order to better reproduce the spatial distribution of jets from newly forming stars. 
This has the advantage of allowing \torch\ to be used to explore the interplay between different feedback mechanisms, the impact of jets from different portions of the IMF, and other questions dependent on the stellar population.

Future development of \torch\ could include implementation of pre-main sequence evolutionary tracks for star particles in order to link protostellar jets to the appropriate stellar evolutionary stage.
Implementing pre-main sequence tracks in \torch\ will require additional code development and changes to the base star formation routine.
Indeed, \cite{Wilhelm2023} discuss potential strategies for implementing pre-main sequence evolution in \torch\ and the potential impact of pre-main sequence evolution and feedback on protoplanetary disks.
Since jets and disks form together, exploring protostellar jets and disks together in \torch\ is also a promising area for future investigation.

\subsubsection{Jet Injection Rate}

The lack of pre-main sequence tracks also impacts the injection rate and lifetime of the protostellar jet.
We choose a constant jet velocity and mass injection rate for each star particle. 
However, protostellar jets are driven by the interaction of the accretion disk with the newly forming star \citep[see, e.g.,][]{Shu+1988, Pelletier+1992, bontemps96, LyndenBell2003, Frank2014, Bally2016, Kolligan2018, Rosen+2020}.
Observations show that jets are bursty and that the average injection rate depends on the accretion rate \citep[e.g.,][]{Shepherd+1996, Reipurth2002, Raga2002, Bally2016, Nisini2018, Rubinstein2023a, Chauhan2024,Omura2024}.
Our approximation of a constant jet velocity and mass injection rate could be relaxed by introducing time variation of the jet injection based on a subgrid model for the accretion rate and star-disk interactions, for instance by using pre-main sequence tracks.
Similarly, the jet lifetime could be tied to the duration of the pre-main sequence evolutionary stage, either by using a sub-grid model for the jet lifetime or by implementing pre-main sequence tracks.

\subsection{X-ray Emission} \label{sec:xrays}

We find that the gas injected by the jets reaches temperatures of up to $10^6$~K in our models. 
Using both observations and simulations, \cite{Lopez-Santiago2015} show that protostellar jets can produce plasma up to $10^7$~K, consistent with X-ray emission.
A variety of observations have indeed detected X-ray emission from protostellar jets \citep[e.g.,][]{Grosso2006,Bonito2010,Grosso2020}.
Thus, the temperatures reached in our models agree with observed properties of protostellar jets.

\subsection{Randomness and the Star Formation Prescription} \label{sec:random}

There is sensitivity to randomness within the \torch\ simulation framework.
Although we have fixed the random seed for most of the random numbers in the code, there are cascading effects from the remaining random variations in the code.

An important impact of randomness in our simulations is the use of a random seed when sampling stellar masses.
This also means that whether a given run forms a high mass star (such as in \snojettwo) is a consequence of the star formation prescription, as described in Section~\ref{sec:star_formation}.
Because the lists of stellar masses are randomly sampled as each sink forms, any sink can form a high mass star if it can accrete enough mass.
This accounts for some of the pauses in star formation where a run will form relatively few star particles for a time before suddenly forming many more.
This can happen if a sink particle is waiting for enough mass to form the next star.
This effect becomes more significant in smaller mass clouds that have fewer sink particles.

When each sink particle forms, it generates a full list of star masses that is sampled from a Kroupa IMF, and stars are only formed in the order of that list as sufficient mass is accreted.
This means that whether a single sink particle accretes a slightly different amount of gas can determine whether or not the run can form a high mass star.
Although we set the same random seed for the sampling of the star mass lists for sink particles, there are other sources of randomness in our runs, including the rotation of the rays in the radiative transfer method and numerical noise that contribute to slightly different gas evolution between runs.
Even a slightly different evolution can lead to different numbers of sinks forming, each with a newly generated random list, thus potentially dramatically altering which star masses actually form.

Our lower mass clouds are particularly sensitive to this randomness in the star formation prescription due to the relatively low number of sink particles that are formed compared to the \verb|M| clouds discussed in the next section. 
Given a smaller number of sink particles, the presence or absence of a particular sink will have a proportionally larger impact on the overall stellar population.

This setup also allows even low-mass clouds to form very high mass stars.
This effect is evident in the \snojettwo\ and \sdeftwo\ simulations, which both formed a $>50~\msun$ star despite having an initial mass of only $5 \times 10^3~\msun$, and in the \mdef\ simulation run, which formed a $>100~\msun$ star despite an initial cloud mass of $2 \times 10^4~\msun$.
It is likely unrealistic for clouds of these sizes to form such massive stars \citep[e.g.,][]{Grudic+2023}.
Indeed, \cite{Weidner2006} and \cite{Weidner2010} show that the highest mass star in a star cluster is tied to the mass of the cloud.
Furthermore, as seen in in our results here (e.g., Figures~\ref{fig:sm_sfe}~and~\ref{fig:med_stars}) and as described by \cite{Lewis+2023}, high mass stars dramatically alter the subsequent evolution of the run and the star clusters that are formed.
Given the fact that our star formation prescription uniformly samples the IMF, this behavior could be limited by altering the upper limit of the range over which the IMF is sampled, for example following \cite{Weidner2013a}.
For the simulations presented here, though, we chose to use the same limit for each run.

\section{Conclusions} \label{sec:conclusions}

In this paper, we present a new module that expands the \torch\ framework for modeling star formation to include protostellar jets as an additional form of stellar feedback. 
Our model aligns the protostellar jets with the angular momentum of the stars, which is inherited from the gas accreted to form those stars.
Our jet injection region is adapted from the model described in \cite{Cunningham+2011} and has both angular and radial dependence.
We introduce how the jets module functions and present results from the first suite of \torch\ runs to include protostellar jets. 
We find that:
\begin{enumerate}
    \item The inclusion of protostellar jets slows and suppresses star formation even in star-forming regions with masses up to $2 \times 10^{4}~\msun$. The \mdef~and \mshort~runs (with jets with default and shortened lifetimes) reach a lower integrated SFE than the \mnojet~run (without jets) in the same amount of time (a difference of more than 15\% by 3~\tff).
    \item The chosen jet parameters, including the lifetime and injection velocity of the jet, significantly impact the effectiveness of protostellar jets at impeding star formation. For example, the integrated SFEs of simulations with different jet parameters in our lower mass clouds differ by as much as 19\% at later times. 
    \item Jets with higher injection velocities are more effective at slowing star formation: the \sslow~run (with jets with a lower injection velocity) reaches a higher integrated SFE than the \sfast~run (with jets with a higher injection velocity) at the same time. We hypothesize that this is due to the increased momentum of the injected material in the \sfast~run. 
    \item The inclusion of protostellar jets alters the energetics of our simulations, including in clouds of up to $2 \times 10^{4}~\msun$. The runs with jets showed an increase in the kinetic energy at early times relative to runs without jets.  We hypothesize that this is the main mechanism by which jets reduce the SFR.
\end{enumerate}

This new module for the inclusion of protostellar jets enables a wide variety of future work exploring the role of protostellar jets in the formation and evolution of stellar clusters. 
Future investigations can now use \torch\ to explore the impact of protostellar jets on the development and evolution of the stellar cluster properties within star-forming regions, the role of protostellar jets in higher mass clusters, and much more.

\begin{acknowledgments}

A draft of this paper (without Section~\ref{sec:energy}, the appendices, and various edits throughout) was included as Chapter 4 of the Ph.D. thesis of S.M.A \citep{Appel2024}. 
S.M.A. would like to thank her thesis committee for their insights on the thesis, which included feedback that significantly improved this paper.

S.M.A. thanks the National Science Foundation (NSF) for support through an Astronomy and Astrophysics Postdoctoral Fellowship; this material is based upon work supported by the NSF under Award No. 24-01740.
S.M.A. \& B.B. acknowledge support from NSF grant AST20-09679.  
M.-M.M.L., E.A., \& B.P. acknowledge support from NSF grant AST23-07950. 
E.A. \& M.-M.M.L. acknowledge support from NASA ATP grant 80NSSC24K0935. 
C.C.-C. is supported by a Canada Graduate Scholarship -- Doctoral (CGS D) from NSERC, and also acknowledges previous support from the Government of Ontario through a Queen Elizabeth II Graduate Scholarship in Science and Technology.
B.B. acknowledges support from NSF grant AST24-07877, the David and Lucile Packard Foundation, the Alfred P. Sloan Foundation,  
and the Mathematics and Physical Sciences (MPS) division of the Simons Foundation. 
 M.J.C.W. is supported by NOVA under project number 10.2.5.12.
A.T. acknowledges support from the Department of Energy Fusion Energy Sciences Postdoctoral Research Program, administered by the Oak Ridge Institute for Science and Education (ORISE) and Oak Ridge Associated Universities (ORAU) under DOE contract DE-SC0014664; and prior support through a NASA Cooperative Agreement awarded to the New York Space Grant Consortium.

The authors thank Joe Glaser, Lourens Veen, Steven Rieder, Shyam Menon, Jonathan Tischio, Ulrich Steinwandel, and the Flatiron Institute's Scientific Computing Core for conversations about \verb|FLASH|, \verb|AMUSE|, and \verb|MPI| that supported the development of this work. 

The authors acknowledge the Simons Foundation for the computational resources used in this work, the Flatiron Institute and the American Museum of Natural History for hosting workshops that facilitated progress on this code, and the Interstellar Institute's programs ``With Two Eyes" in 2022 and ``ii7'' in 2025 at the Paris-Saclay University's Institut Pascal for hosting discussions that nourished the development of the ideas behind this work.

The software used in this work was developed in part by the DOE NNSA- and DOE Office of Science-supported Flash Center for Computational Science at the University of Chicago and the University of Rochester.

\end{acknowledgments}

\begin{contribution}
S.M.A. led the code development, performed and analyzed the simulations used, wrote the manuscript, and led the editing and submission of the manuscript.

B.B. advised the Ph.D. thesis of S.M.A. of which this paper formed a part, including providing funding that contributed to S.M.A.'s thesis work.

B.B. and M.-M.M.L. supervised this project, including contributing to the code design, providing input on the plan for and running the simulation suite, guiding the analysis, and giving feedback on the manuscript. 

All subsequent authors are listed alphabetically. 

E.P.A. contributed to the code design, provided input on the plan for and running the simulation suite, guided the analysis, and gave feedback on the manuscript.

C.C.-C. contributed to the code design, provided input on the plan for and running the simulation suite, co-designed the initial conditions for the simulation suite, guided the analysis, and gave feedback on the manuscript.

S.L. contributed to the code design, provided input on the plan for and running the simulation suite, and gave feedback on the manuscript.

S.L.W.M. contributed to the code design and provided input on the plan for and running the simulation suite.

B.P. contributed to the code design, provided input on the plan for and running the simulation suite, guided the analysis, and gave feedback on the manuscript.

S.P.Z. contributed to the code design, provided input on the plan for and running the simulation suite, guided the analysis, and gave feedback on the manuscript.

A.T. contributed to the code design and gave feedback on the manuscript.

M.J.C.W. contributed to the code design and gave feedback on the manuscript.

\end{contribution}

\software{ 
\texttt{AMUSE} \citep{PortegiesZwart2009, Pelupessy+2013, PortegiesZwart2013, PortegiesZwartMcMillan2018}, \texttt{AMUSE-SeBa} \citep{PortegiesZwartVerbunt1996,Toonen+2012}, \texttt{AMUSE-ph4} \citep{McMillan+2012, PortegiesZwartMcMillan2018},
 \texttt{yt} \citep{Turk11a}, \texttt{FLASH} \citep{Fryxell2000,dubey2014}, \texttt{SciPy} \citep{SciPy}, \texttt{Matplotlib} \citep{Hunter2007}, \texttt{astropy} \citep{Astropy-Collaboration13a} 
}

\begin{appendix}

\section{Single Jet Resolution Study} \label{sec:single_jet_resolution}

\begin{figure*}[h!tb]
\centering
\includegraphics[width = 0.7\linewidth]{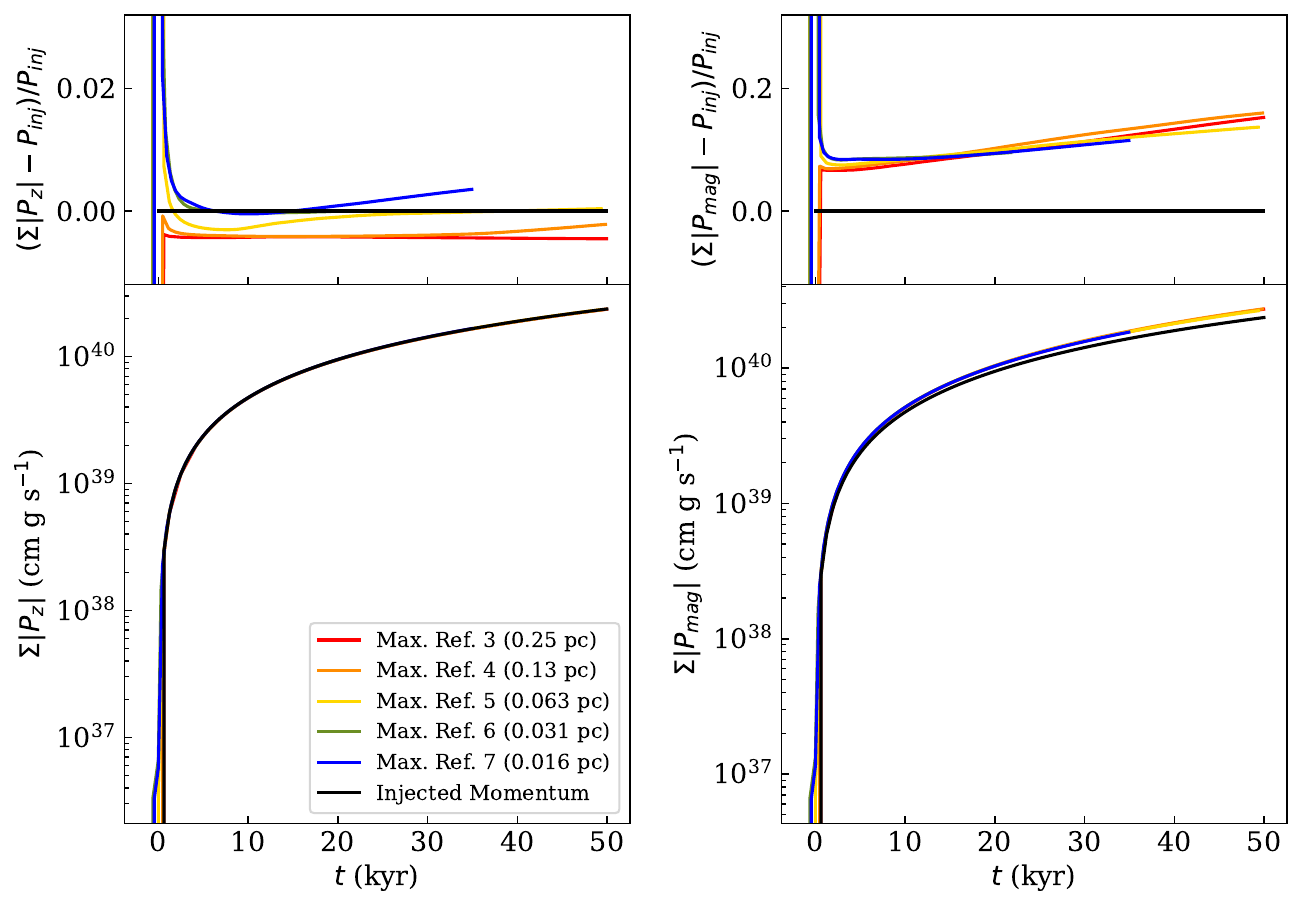}
\caption{ A comparison of the analytic expectation of the total injected momentum (black line; based on the default jet parameters and a 5~$\msun$ star particle) to the actual total z-direction momentum of the gas (left column) and the total momentum magnitude of the gas (right column) as functions of time since a single jet began to be injected. The lower panels show the momentum values, while the upper panels show the percent differences relative to the expected total injected momentum. Runs with 5 different maximum refinement levels are shown and the corresponding minimum cell size is listed in the legend.
\label{fig:res_momentum}}
\end{figure*}

We present five runs with a box size $L=16$~pc and five different maximum refinements.
Each run injects a jet from a single star particle into a uniform density background medium of $2.18 \times 10^{-23}$~g~cm$^{-3}$.
The highest resolution of these runs is presented in Figure~\ref{fig:jet_rendering} above.
Each of these jets are injected by a 5.0~$\msun$ star particle using the default jet parameters described in Table~\ref{tab:user_params} and have minimum cell sizes of 0.250, 0.125, 0.0625, 0.0313, and 0.0156~pc.
The two highest refinement runs do not progress as long as the lower refinement runs.
These runs demonstrate the evolution of a jet injected with our model when in isolation (i.e., no other feedback mechanisms or variations in gas density) and how the jet injection varies with resolution.

Figure~\ref{fig:res_momentum} shows the total z-direction momentum and total momentum magnitude for each of these single star tests as functions of time.
The black lines indicate the analytical expected momentum injected by the jet, based on the $dm/dt$ and injection velocity for a jet from a 5~$\msun$ star particle and the default jet parameters.
The bottom panels of Figure~\ref{fig:res_momentum} show the actual momentum values, while the top panels show the percent difference between the actual momentum on the grid and the expected total injected momentum.

Overall, the difference between the injected momentum (the black lines in Figure~\ref{fig:res_momentum}) and the actual z-direction momentum on the grid is small.
There is a significant discrepancy initially as the injection region is established, but after a few kyrs the difference between the analytically expected injected momentum and the z-direction momentum is less than 1\% for all five runs at early ($\sim8$~kyr) and late times (50~kyr for the three lower resolution runs, 22~kyr for the second highest resolution run, and 35~kyr for the highest resolution run).

The difference between the analytically expected injected momentum (the black lines in Figure~\ref{fig:res_momentum}) and the total momentum on the grid is less than 9\% for all five runs at early times, and less than 16\% at late times.
We expect that the reason for the larger discrepancy between the total momentum and the injected momentum is the contribution to the momentum by expanding hot gas.
This also explains the growing difference at later times for both the total momentum and the z-direction momentum as the amount of expanding hot gas on the grid grows.

The runs with different maximum refinement levels follow the same behavior and differ by less than a few percent from each other.
The minimum cell size (0.137~pc) for the runs presented in the main body of this paper fall between the two lowest refinement single-jet runs in this section (0.250~pc and 0.125~pc).
Since these two runs behave similarly to the highest resolution single-jet runs and inject a very similar amount of momentum, we we do not expect that increasing the resolution of our star-forming cloud runs will significantly alter the impact of jet feedback on the energetics and overall properties of our star-forming clouds.

\section{Cloud Resolution Study} \label{sec:jets_resolution}

\begin{deluxetable*}{lcccccccr}[thb!]
\tabletypesize{\footnotesize}
\tablecaption{Summary of the runs used for the cloud resolution study. This test suite includes runs with three different maximum refinements but otherwise identical setups. The highest resolution run \shi\ does not reach a full 3~\tff\ and so the SFE and $N_{\mathrm{*}}$ for each run is shown at $t=5.5$~Myr which corresponds to 2.5~\tff.}
\label{tab:resolution_study}
\tablecolumns{5}
\tablewidth{0pt}
\tablehead{
\colhead{Sim.\ Name} & 
\colhead{$M_{\mathrm{cloud}}$ ($\msun$)} & 
\colhead{$\Sigma_{\mathrm{init}}$ ($\msun \mathrm{pc}^{-2}$)} & 
\colhead{Min.\ Cell (pc)} & 
\colhead{Jets} & 
\colhead{Jet Time (kyr)} & 
\colhead{Jet Vel.} &
\colhead{SFE$_{5.5}$ (\%)} & 
\colhead{$N_{\mathrm{*},5.5}$}
}
\startdata
    \texttt{S-jet}$^{\dagger}$ & $5 \times 10^{3}$  & $32.5$  & $0.137$   & On  & $100$ & 0.25  & $14.4 $ & $1,151$ \\
    \texttt{S-jet2}$^{\dagger}$ & $5 \times 10^{3}$  & $32.5$  & $0.137$   & On  & $100$ & 0.25  & $7.44 $ & $482$ \\
    \texttt{S-jet3}$^{\dagger}$ & $5 \times 10^{3}$  & $32.5$  & $0.137$   & On  & $100$ & 0.25  & $11.1 $ & $973$ \\
    \texttt{S-lowres}  & $5 \times 10^{3}$  & $32.5$  & $0.273$   & On  & $100$ & 0.25  & $13.9 $ & $854$ \\
    \texttt{S-hires}   & $5 \times 10^{3}$  & $32.5$  & $0.0684$ & On  & $100$ & 0.25  & $8.90 $ & $688$ 
\enddata
\tablecomments{\textit{Sim.\ Name}: name of simulation used in the text;  $M_{\mathrm{cloud}}$: initial total mass of the spherical cloud; $\Sigma_{\mathrm{init}}$: initial surface density at center of the sphere; \textit{Min.\ Cell}: minimum cell size; \textit{Jets}: whether protostellar jets are turned on; \textit{Jet Time}: value of \texttt{jet\_time}; \textit{Jet Vel.}: value of \texttt{jet\_vel\_fraction}; \textit{SFE$_{5.5}$}: integrated star formation efficiency after 2.5~\tff; $N_{\mathrm{*},5.5}$: the total number of stars formed after 2.5~\tff. $^{\dagger}$These runs are also shown in Table~\ref{tab:testruns} and are reproduced here for comparison.}
\end{deluxetable*}

\begin{figure*}[h!tb]
\centering
\includegraphics[width = 0.95\linewidth]{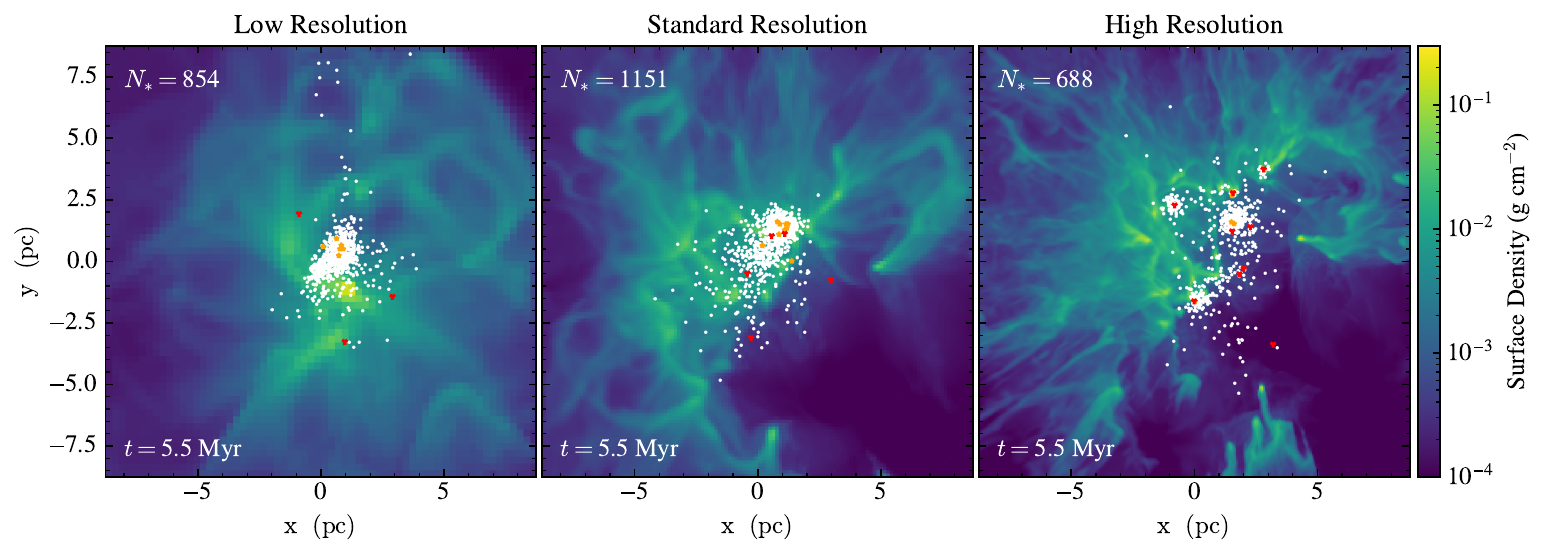}
\caption{Projection plots of density along the z-axis are shown for three runs with varying maximum refinements: \slo, \sdef, and \shi.
All runs include jets with the default values for the jet parameters, as in \sdef.  
The x-y position of each star particle is shown as a white dot for non-feedback stars and an orange star for any star injecting either winds or jets. Sink particles are shown as red Y symbols. The number of star particles is shown on each panel. For these runs $t_{\mathrm{ff}} = 2.2$~Myr, so the plotted time corresponds to approximately 2.5~\tff.
\label{fig:res_projections}}
\end{figure*}

\begin{figure}[h!tb]
\centering
\includegraphics[width = \linewidth]{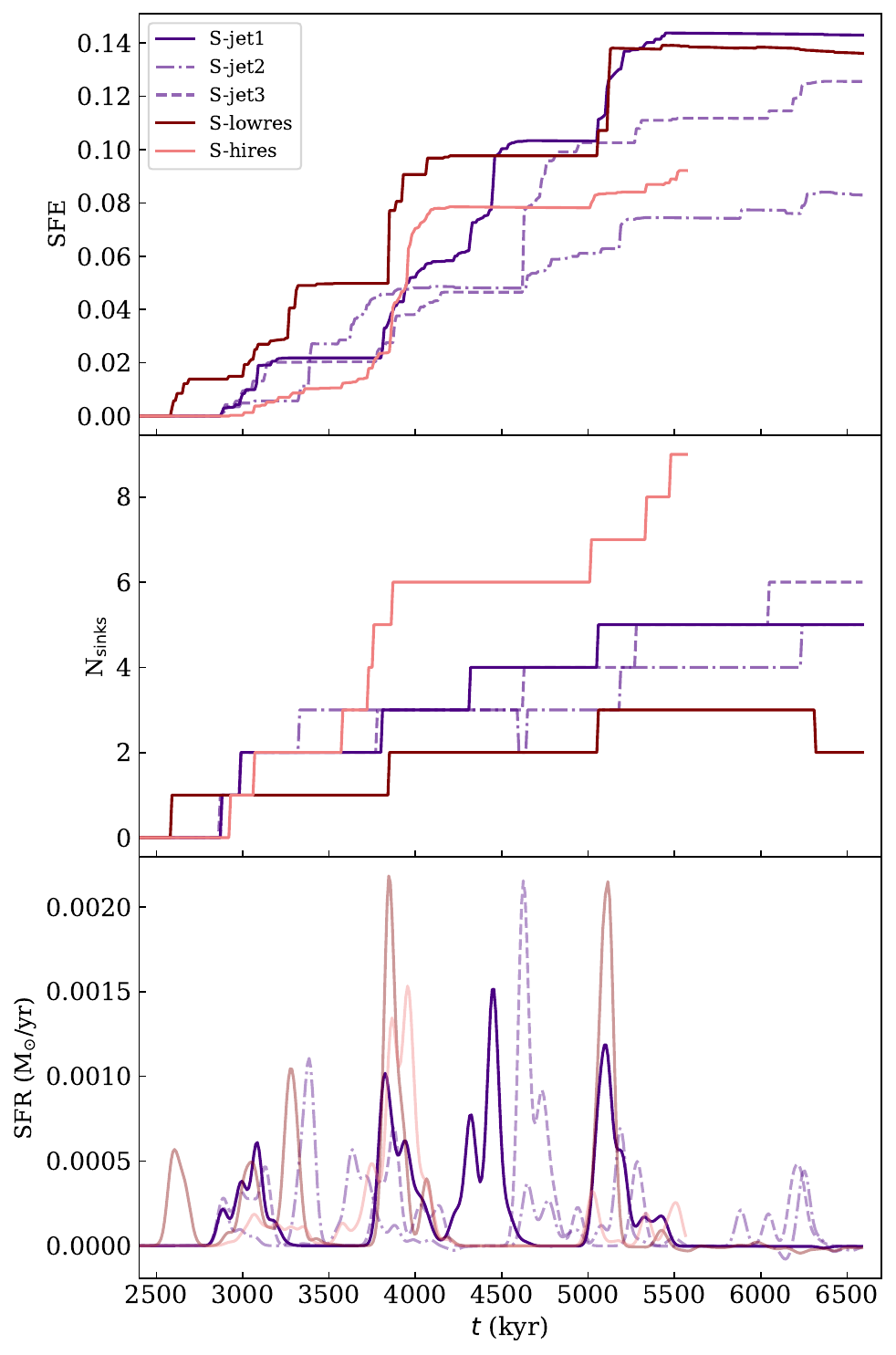}
\caption{ A comparison of the integrated SFE, the number of sink particles, and the smoothed SFR as functions of time for the five runs shown in Table~\ref{tab:resolution_study}. 
The bottom panel shows the SFR smoothed with a Gaussian filter with a sigma of 3, which corresponds to approximately 30~kyr.
\label{fig:res_sfe}}
\end{figure}

In addition to the runs described in Section~\ref{sec:testing} (see Table~\ref{tab:testruns}), which explore different jet parameters and cloud masses, we considered two additional runs that explore the effect of the maximum resolution of the simulation. 
These new runs are listed in Table~\ref{tab:resolution_study} alongside the three fiducial jets runs from Table~\ref{tab:testruns}.
The \slo\ and \shi\ runs use the same set up as runs \sdef, \sdeftwo, and \sdefthree\ except that they have lower and higher maximum refinements, respectively.
The corresponding minimum cell sizes are shown in Table~\ref{tab:resolution_study}.
Due to the significantly higher computational cost of the \shi\ run, it did not progress to a full 3~\tff\ and Table~\ref{tab:resolution_study} instead reports the SFE and $N_{\mathrm{*}}$ as of 2.5~\tff\ (5.5~Myr) for each run.
Figure~\ref{fig:res_sfe} shows the integrated SFE, the number of sink particles, and the smoothed SFR for each of these runs in comparison to \sdef, \sdeftwo, and \sdefthree.
The SFE and SFR quantities are comparable Figures~\ref{fig:sm_sfe}~and~\ref{fig:med_sfe}.
The number of sink particles is comparable to the second panel in Figures~\ref{fig:sm_stars}~and~\ref{fig:med_stars}.

The overall progress of star formation, as measured by the SFE, is similar for each of the runs (see Figure~\ref{fig:res_sfe}).
The higher resolution runs are slightly slower to start forming stars than the lowest resolution runs.
The \shi\ run shows the same overall behavior as the \sdef, \sdeftwo, and \sdefthree\ runs, with differences that are comparable to the differences between the three identical runs.
The central panel of Figure~\ref{fig:res_sfe} shows that the higher resolution runs form a much larger number of sink particles, with \shi\ forming as many as 9 sink particles and \slo\ never forming more than 3 sink particles.

Overall, we find that the resolution chosen for the majority of our runs is sufficient, given that the overall behavior of the \sdef, \sdeftwo, and \sdefthree\ runs is comparable to the behavior of the \shi\ run and considering the significant increase in computational cost of the \shi\ run.
Furthermore, Appendix~\ref{sec:single_jet_resolution} shows that the momentum injected by individual jets does not significantly alter with resolution higher than that of the majority of our runs, suggesting that increasing the resolution will not substantially alter the role of protostellar jets in our runs.

\section{Injection Region Model Comparison} \label{sec:dtheta}

\begin{figure}[b!]
\centering
\includegraphics[width = \linewidth]{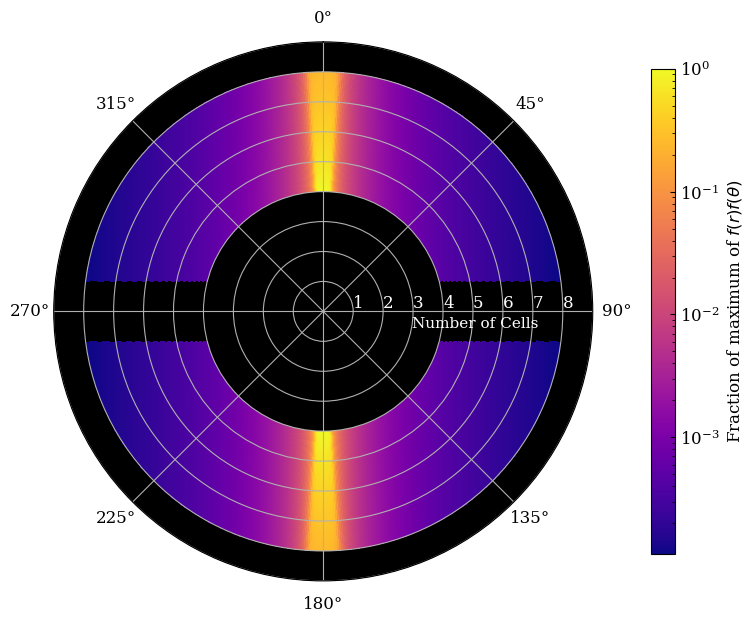}
\caption{Figure~\ref{fig:inj_region} but with a fixed $\Delta \theta = \arctan(1/8)$ as in \cite{Cunningham+2011}. This is the injection region produced by the version of the \torch\ code identified by the \texttt{jets-v1.2} tag or newer and is recommended for future studies using \torch\ with jets. The heatmap shows the relative value (as a fraction of the maximum value) of the product of the radial and angular dependence: $f(r)g(\theta)$. The actual weighting for each cell of the injection region is the product of this quantity evaluated at the cell center and the geometric factor determining the overlap of the injection region and the specific cell, and is normalized such that all the weights add to 1. Neither this normalization nor the geometric factor are included in this plot, in order to focus on the distribution of the $f(r)g(\theta)$ factor. Regions shaded black are outside the injection region and have a weighting of 0. 
\label{fig:inj_region_fixeddtheta}}
\end{figure}

In this appendix we test the impact of a variable $\Delta \theta$ (as in the runs presented so far) versus a fixed $\Delta \theta$ \cite[as in][]{Cunningham+2011}.
We performed three new single jet runs using the version of the Torch code at commit \href{https://bitbucket.org/torch-sf/torch/commits/1aceb49afc8e332264100b0ca51b361774fac294}{1aceb49} and tagged as \verb|jets-v1.2| in the \torch\ repository. 
The only change to the code for this test was to implement a fixed $\Delta \theta = \arctan(1/8)$ in Equations~\ref{eq:ftheta}~and~\ref{eq:psi} instead of $\Delta\theta = \Delta_{\mathrm{min}} / r$, as is described in Section~\ref{sec:rad_ang_dep}. 
This change brings the code into closer alignment with the model described in \cite{Cunningham+2011}.
In the \cite{Cunningham+2011} model, $\Delta \theta$ is used when smoothing the function for the angular dependence over an angle subtended by a cell at the outer edge of the injection region. 
In contrast, setting $\Delta\theta = \Delta_{\mathrm{min}} / r$ adapts the window of this averaging step to the angle subtended by a cell at radius $r$, which results in an additional radial dependence in the final expression for the angular dependence relative to a fixed $\Delta \theta$ value.

We note that because the jets model we implement in \torch\ normalizes the strength of our injected material over the whole injection region after using $\Delta \theta$ to calculate the injection weighting (Equation~\ref{eq:weighting}), the change to $\Delta \theta$ does not alter the total amount of mass injected or the injection velocity.  
Instead, $\Delta \theta$ results in a small alteration to the shape of the injection region similar to altering the collimation.
Figure~\ref{fig:inj_region_fixeddtheta} shows the same map of the injection region from Figure~\ref{fig:inj_region} for $\Delta \theta = \arctan(1/8)$.
The injection regions are similar, but with a more pronounced radial dependence in Figure~\ref{fig:inj_region_fixeddtheta}.

Our three new test runs have different maximum refinements, corresponding to minimum cell sizes of 0.25, 0.125, and 0.0625~pc. 
Aside from the change to the value of $\Delta \theta$, these runs are identical in setup to the single jet runs in Appendix~\ref{sec:single_jet_resolution} with maximum refinement levels 3, 4, and 5.
Examining projection plots of each run by eye, we see no visible difference in the gas morphology between runs with each value of $\Delta \theta$, for a given refinement level.
We also considered the total expected injected momentum versus the actual z-direction momentum and momentum magnitude (analogous to the quantities in Figure~\ref{fig:res_momentum}).
After 50~kyr, there is a $1.1\%$ difference in the amount of momentum injected on the grid in the direction of the jet between the two models for the runs with maximum refinement levels 4 and 5, and a $1.2\%$ difference for the runs with maximum refinement level 3.  
The difference in the total magnitude of the momentum on the grid between the two models after 50~kyr ranges between $0.01\%$ and $0.7\%$, depending on refinement level.
We conclude that the change in the value of $\Delta \theta$ has functionally no impact on our runs and that we have no reason to think that this change to the model alters our results for full clouds.

Future versions of the jets code with \torch\ will implement the $\Delta \theta$ value in \cite{Cunningham+2011} for the sake of consistency with previous work.  However, we do not anticipate that this change will alter future results.

\end{appendix}

\bibliographystyle{aasjournal}
\bibliography{my_bib.bib}

\end{document}